\documentclass[12pt,preprint]{aastex}

\usepackage{graphicx}
\usepackage{amssymb}
\usepackage{multirow}

\shorttitle{Spectral hardness evolutionary slope of FRED pulses}
\shortauthors{Peng et al.}

\begin{document}


\title{The $E_{p}$ Evolutionary Slope within the Decay Phase of ``FRED'' Gamma-ray Burst Pulses}

\author{Z. Y. Peng \altaffilmark{1}, L. Ma \altaffilmark{1,{\ast}}, X. H. Zhao \altaffilmark{2}, Y. Yin\altaffilmark{1}, L. M. Fang\altaffilmark{3}, Y. Y.
Bao\altaffilmark{4}}

\altaffiltext{1}{Department of Physics, Yunnan Normal University,
Kunming 650092, China; pzy@ynao.ac.cn}

\altaffiltext{${\ast}$}{Corresponding author, astromali@126.com }

\altaffiltext{2}{Department of Astronomy, Nanjing University,
Nanjing, Jiangsu 210093, China}

\altaffiltext{3}{Department of Physics, Guangdong Institute of
Education, Guangzhou 510303, China}

\altaffiltext{4}{Department of Physics, Yuxi Normal College, Yuxi
653100, China}


\begin{abstract}

Employing two samples containing of 56 and 59 well-separated FRED
(fast rise and exponential decay) gamma-ray burst (GRB) pulses whose
spectra are fitted by the Band spectrum and Compton model,
respectively, we have investigated the evolutionary slope of $E_{p}$
(where $E_{p}$ is the peak energy in the $\nu F\nu$ spectrum) with
time during the pulse decay phase. The bursts in the samples were
observed by the Burst and Transient Source Experiment (BATSE) on the
Compton Gamma-Ray Observatory. We first test the $E_{p}$
evolutionary slope during the pulse decay phase predicted by Lu et
al. (2007) based on the model of highly symmetric expanding
fireballs in which the curvature effect of the expanding fireball
surface is the key factor concerned. It is found that the
evolutionary slopes are normally distributed for both samples and
concentrated around the values of 0.73 and 0.76 for Band and Compton
model, respectively, which is in good agreement with the theoretical
expectation of Lu et al. (2007). However, the inconsistence with
their results is that the intrinsic spectra of most of bursts may
bear the Comptonized or thermal synchrotron spectrum, rather than
the Band spectrum. The relationships between the evolutionary slope
and the spectral parameters are also checked. We show the slope is
correlated with $E_{p}$ of time-integrated spectra as well as the
photon flux but anticorrelated with the lower energy index $\alpha$.
In addition, a correlation between the slope and the intrinsic
$E_{p}$ derived by using the pseudo-redshift is also identified. The
mechanisms of these correlations are unclear currently and the
theoretical interpretations are required.

\end{abstract}

\section{INTRODUCTION}
The origin of the gamma-ray burst (GRB) still remains unclear though
it has been found for more than forty years and many progresses have
been made. The best probe of early phase of jet attributes and the
physical mechanism is the prompt emission, essentially the
temporal-spectral dependence of GRB pulses due to the absence of
some other ways to detect them, such as gravitational wave and
neutrino detections. The analysis of the GRB prompt emission
provides us valuable clues to the environment from which the
radiation emits and the underlying processes giving rise to the
phenomenon.
The spectral evolution is universal and has been studied both over
the entire burst, giving the overall behavior and over individual
pulse structures (see, for instance, the review by Ryde 1999).
Pulses are common features in a GRB light curve and appear to be the
fundamental constituent of it (see, e.g., Norris et al. 1996; Stern
\& Svensson 1996). This single pulse temporal evolution is often
been described by a fast rise and exponential decay (the so-called
FRED shape, see Fishman et al. 1994). The spectral hardness
decreases during pulse decay and there are two empirical relations
between the temporal and spectral properties to characterize the
spectral evolution. One important correlation is that between
hardness of the spectrum, the peak energy, and fluence (HFC; Liang
\& Kargatis 1996). The other correlation is that between hardness
and the instantaneous flux (or intensity) (HIC; Golenetskii et al.
1983). The two relations appear to be satisfied by a substantial
fraction of GRB pulses during the decay phase (see, Crider \& Liang
1999; Ryde \& Svensson 2000). Most studies concerning these
correlations examine them in single pulses and do not compare the
behavior of pulses within a burst. Combining the two correlations,
i.e., the HIC and HFC, Ryde \& Svensson (2000) showed that a
power-law HIC and an exponential HFC resulted in the decay phase of
the pulse following power-law behaviors described by equations 3 and
4 in their paper. Ryde \& Svensson (2002) studied a sample
containing 25 pulses and found that a power law gives a better
description of the pulse decays than a stretched exponential, the
most commonly assumed pulse shape so far. They also found that there
are no obviously preferred values of the power-law index of peak
energy, $E_{p}$, in the decay phase of pulse.

Daigne \& Mochkovitch (2003) presented a simple, semi-analytical
model to interpret the GRB temporal and spectral properties in the
context of the internal shock model. The spectral evolution of
synthetic pulses was first obtained with standard equipartition
assumptions to estimate the post-shock magnetic field and the
electron Lorentz factor. They found that $E_{p}$ $\propto $
$t^{-\delta}$, which shows the decay of peak energy during the
pulses decay phase follows the power law relation with time. They
also compared the consistence of the power-law index, $\delta$, of
their model with the observed index provided by Ryde \& Svensson
(2002) and found that the synchrotron process with standard
equipartition assumptions gives a much too steep spectral evolution,
which showed that another process different from synchrotron one
might radiate that energy.

The observed GRB pulses are believed to be produced in a
relativistically expanding and collimated fireball because of the
large energies and the short time-scales involved. The so-called
Doppler effect (or curvature effect in some paper), which over the
whole fireball surface would play an important role to account for
the observed pulses and spectra of not only prompt GRBs but the
early X-ray afterglow, is that the photons emitted from the regions
on the line of sight and off the line of sight are Doppler-boosted
by different factors and travel different distances before reaching
the observer (e.g. Meszaros and Rees 1998; Hailey et al. 1999; Qin
2002, Qin et al. 2004, Qin 2008a, 2008b, 2008c). The curvature
effect model of a cutoff power law spectrum has also been used to
model the light curve and spectral evolution of the X-ray tail
(Zhang et al. 2009). Based on the model of Doppler effect Lu et al.
(2007) (hereafter Paper I) investigated the evolution of observed
spectral peak energy $E_{p}$ and found that the evolutionary curve
of $E_{p}$ undergoes a drop-to-rise-to-decay phase. The decay phase
of the pulse, where the Doppler effect dominates, always decreases
monotonically and does not necessarily reflect the corresponding
intrinsic spectral evolution. They first explored the case the
intrinsic spectrum is the Band function (Band et al. 1993) for three
different local pulses. The three decay phases of $E_{p}$ within the
decay phase of the light curve were extracted and performed a linear
least-squares fit to the $E_{p}$ and relative observed time, $\tau$,
and have $\log E_{p} = I - S \log \tau$, with the same slope of $S =
0.95 \pm 0.01$ for the three curves, which indicated that the slope
of the decay phase of $E_{p}$ is not affected by the shape of its
local pulse. Then they studied the case of intrinsic Comptonized and
thermal synchrotron spectrum and found that corresponding slope $S =
0.75 \pm 0.01$ for both of the two intrinsic spectra, which also
showed that the slope is independent of the forms of local pulse.

Although several GRB missions such as HETE-2 and Swift have been
launched, the Burst and Transient Source Experiment (BATSE; Fishman
et al. 1989), aboard the Compton Gamma Ray Observatory (CGRO;
Gehrels et al. 1994), provided the largest GRB database from a
single experiment among all the gamma-ray experiments that have
detected GRBs. The BATSE data are still the most suitable for
detailed spectral studies of GRB prompt emission, both in quantity
and quality. For many of the BATSE GRBs, high time and energy
resolution data are available. BATSE also provided wider energy
coverage than current GRB missions.

In the present work, we select two samples observed by BATSE  to
investigate the evolutionary slope of $E_{p}$ during the decay phase
of GRB pulses. Different from what Ryde \& Svensson (2002) did, we
adopt those well-separated FRED pulses compiled by Peng et al.
(2007). First of all, we check whether the observed evolutionary
slope of $E_{p}$ during the pulse decay phase is consistent with the
theoretical predication of Paper I. Examining whether the slope of
$E_{p}$ during the decay phase of pulse is related to other
parameters such as $E_{p}$ of time-integrated spectra is another
motivation of this work. In section 2, we present the sample
description and spectral modeling. The results are given in section
3. Discussion and conclusions are presented in the last section.


\section{SAMPLE DESCRIPTION AND SPECTRAL MODELING}

In this work we only select the GRB sample compiled by Peng et al.
(2007) consisting of two samples provided by Kocevski et al. (2003)
and Norris et al. (1999), in which the bursts are found to contain
individual FRED pulses. The data with durations longer than 2 s  are
provided by the BATSE's LADs instruments on board the CGRO
spacecraft, which provide discriminator rate with 64 ms resolution
from 2.048 s before the burst to several minutes after the trigger
(Fishman et al. 1994) (for more details of the sample selection, see
Kocevski et al. 2003. and Norris et al. 1999). In order to obtain
the evolutionary slope of $E_{p}$ during the pulse decay phase (we
use the symbol S to denote the evolutionary slope of $E_{p}$ during
the pulse decay phase) we must study the time-resolved spectroscopy.
The time-resolved spectral analysis requires data with a high enough
signal-to-noise (S/N) ratio to obtain better statistical results.
Similar to Ryde \& Svensson (2002) and Firmani et al. (2008) we also
choose those bursts whose peak flux, $F_{p}$, above 1.8 photons
$cm^{-2} s^{-1}$ on a 256 ms timescale to ensure a reasonable
determination of the spectral parameters. In this way 78 pulses meet
the criteria.


Before performing the spectral modeling we must select the type of
detector, data type, and time and energy interval. Following the
Kaneko et al. (2006) (hereafter Paper II) we also only use the LAD
data mainly to take advantage of its larger effective area. Three
LAD data types used in this work are Medium Energy Resolution data
(MER), High Energy Resolution Burst data (HERB), and Continuous data
(CONT) (for more detailed description about the characteristics of
each data type one can refer to Paper II). To perform detailed
time-resolved spectroscopy it has been shown by Preece et al. (1998)
that a S/N $\sim$45 is needed. However, a high S/N  will lead to
time-resolved spectra consisting of only a few broader time bins.
Our present work focus on the analysis of the S. Therefore the aim
of our spectral analysis, of every time bin, is mainly to determine
the $E_{p}$ and allowing a deconvolution of the count spectrum to
find the energy spectrum. In order to arrive at reliable results we
need as many time bins as possible to study the S. For this purpose,
we adopt the criterion that the decay phase of the pulses should
have at least five time bins with $S/N \geq 30$ to be included in
the study. We check that the results are consistent with higher S/N
ratios. This gives us the possibility to study the burst pulses with
higher time resolution, especially for the later time bins, which is
of great importance for our study.

Similar to Paper II we also use the spectral analysis software
RMFIT, which was specifically developed for burst data analysis by
the BATSE team (Mallozzi et al. 2005). It incorporates a fitting
algorithm MFIT that employs the forwardfolding method (Briggs 1996),
and the goodness of fit is determined by $\chi^{2}$ minimization.
One advantage of MFIT is that it utilizes model variances instead of
data variances, which enables more accurate fitting even for
low-count data (Ford et al. 1995).

We analyze both time-resolved and time-integrated spectra for each
pulse of our sample. In addition, the entire pulse must be included
in our analysis since these pulses are well-separated. The
time-resolved and time-integrated spectra are modeled with two
photon models. One is the most used so-called Band function (Band et
al. 1993):
\begin{equation} f_{BAND}=A\left\{
\begin{array}{ll}
(\frac{E}{100})^{\alpha}\exp(-\frac{E(2+\alpha)}{E_{peak}})
&  E < E_{c},\\
(\frac{(\alpha-\beta)E_{peak}}{100(2+\alpha)})^{\alpha-\beta}
\exp(\beta-\alpha)(-\frac{E}{100})^{\beta} &  E \geq E_{c},
\end{array}
\right.
\end{equation}
where, $E_{c}=(\alpha-\beta)\frac{E_{peak}}{2+\alpha}\equiv
(\alpha-\beta)E_{0}$. The model consists of four parameters: the
amplitude A in photons s$^{-1}$ cm$^{-2}$ keV$^{-1}$, a low-energy
spectral index $\alpha$, a high energy spectral index $\beta$, and a
$\nu F_{\nu}$ peak energy $E_{peak}$ in keV, which is related to the
e-folding energy, E$_{0}$.

Paper II have showed that the Comptonized Model (COMP model) tends
to be preferable in fitting time-resolved spectra due to better
$\chi^{2}$ values compared with that of BAND model. Additionally,
many BATSE GRB spectra lack high-energy photons (Pendleton et al.
1997). Those no-high-energy spectra are usually fitted well with
this model. This also shows the existence of a number of spectra
without high-energy component. Therefore, the spectra are also
modeled with the COMP model. It is a low-energy power law with an
exponential high-energy cutoff, which is equivalent to the BAND
model without a high-energy power law, namely $\beta \rightarrow
\propto$, and has the form
\begin{equation}
f_{COMP}(E)=A(\frac{E}{E_{piv}})^{\alpha}\exp(-\frac{E(2+\alpha)}{E_{peak}}),
\end{equation}
E$_{piv}$ was always fixed at 100 keV , therefore, the model
consists of three parameters: A, $\alpha$, and $E_{peak}$.

We always chose the data taken with the detector that was closest to
the line of sight to the location of the GRB, as it has the
strongest signal. For the case of weaker bursts the MER data is
preferable because it has much finer time resolution. Whereas for
the bright bursts the HERB data is preferable due to its higher
energy resolution. A background estimate is made using the High
Energy Resolution data (HER) data for the HERB data, covering
several thousand seconds before the trigger and after the HERB
accumulation is finished. In the MER data, the CONT data are used as
background. The light curve of the background, during the outburst,
is modeled with a second- or third-order polynomial fit. The usable
energy range for spectral analysis is $\sim$ 30 keV - $\sim$ 2 MeV
for all the bursts. The lowest seven channels of HERB and two
channels of MER and CONT are usually below the electronic lower
energy cutoff and are excluded. The highest few channels of HERB and
normally the very highest channel of MER and CONT are unbounded
energy overflow channels and also not usable (see, Paper II). The
procedures give us all the spectral parameters modeled by BAND and
COMP model for every time bin of the whole pulses besides energy
flux. Similar to Peng et al. (2009) the following data points are
excluded:

(1) the resulting $\chi^{2}$ per degree of freedom is -1 when
fitting each time-resolved spectrum because in this case the
nonlinear fitting of corresponding time-resolved spectra is failure.

(2) $\alpha < -2$ and $\beta > -2$ for the BAND model, $\alpha < -2$
for the COMP model. As Paper II pointed out that, the fitted
$E_{peak}$ represents the actual peak energy of the $\nu F_{\nu}$
spectrum only in the case of $\alpha > -2$ and $\beta < -2$ for the
BAND model and $\alpha > -2$ for the COMP model.

(3) the uncertainty of its corresponding $E_{peak}$ is larger 50\%
than itself. In this way, we can obtain the best statistic.

With these criterions, we can obtain useful data points of
$E_{peak}$ as well as the other parameters. At the end, we have 56
and 59 pulses for Band model and Comp model, which are denoted with
sample 1 and 2, respectively.

We focus, in this study, our attentions on the decay phase of
pulses. Hence the initial time of the decay phase must be
determined. Similar to Peng et al. (2006, 2009) the
subtract-background pulses, combining the data from the BATSE four
channels are modeled with the function presented in equation (22) of
Kocevski et al. (2003) because we find that this function can well
describe the observed profile of a FRED pulse. In addition, a fifth
parameter, $t_{0}$, which measures the offset between the start of
the pulse and the trigger time, is introduced.
\begin{equation}
F(t)={F_m}(\frac{t+t_0}{t_m+t_0})^r[\frac{d}{d+r}+\frac{r}{d+r}(\frac{t+t_0}{t_m+t_0})^{(r+1)}]^{-\frac{r+d}{r+1}},
\end{equation}
where $t_{m}$ is the time of the pulse's maximum flux, $F_{m}$; r
and d are the power-law rise and decay indexes, respectively. Note
that equation (1) holds for $t\geq -t_0$, when $t< -t_0$ we take
$F(t)=0$.

To obtain an intuitive view of the result of the fit, we develop and
apply an interactive IDL routine for fitting pulses in bursts, which
allows the user to set and adjust the initial pulse parameters
manually before allowing the fitting routine to converge on the
best-fitting model via the reduced $\chi^{2}$ minimization. The fits
are examined many times to ensure that they are indeed the best
ones. The $t_{m}$ of each pulse are found and served as the starting
time of the pulse decay phase. The values, $t_{m}$, $t_{0}$ and
fitting chi-square are listed in Tables 1 and 2. In this way the
data points of $E_{p}$ within the decay portions of the FRED pulses
are extracted.

\section{analysis result}
\subsection{the evolutionary slope seen in the FRED pluses}
Since Paper I have shown that the evolutionary slope in pulse decay
phase can be well modeled with a single power-law form we first
adopt this form to fit the temporal evolution of $E_{p}$ during the
decay phase of pulse.
\begin{equation}
{E_{p}(t)}={\frac{E_{p,0}}{((t+ t_{1})/\tau)^{\delta}}},
\end{equation}

The example plots of the fits within the decay phases are given in
Figures 1 and 2, which correspond to samples 1 and 2, respectively.

\begin{figure}
\centering
\resizebox{3in}{!}{\includegraphics{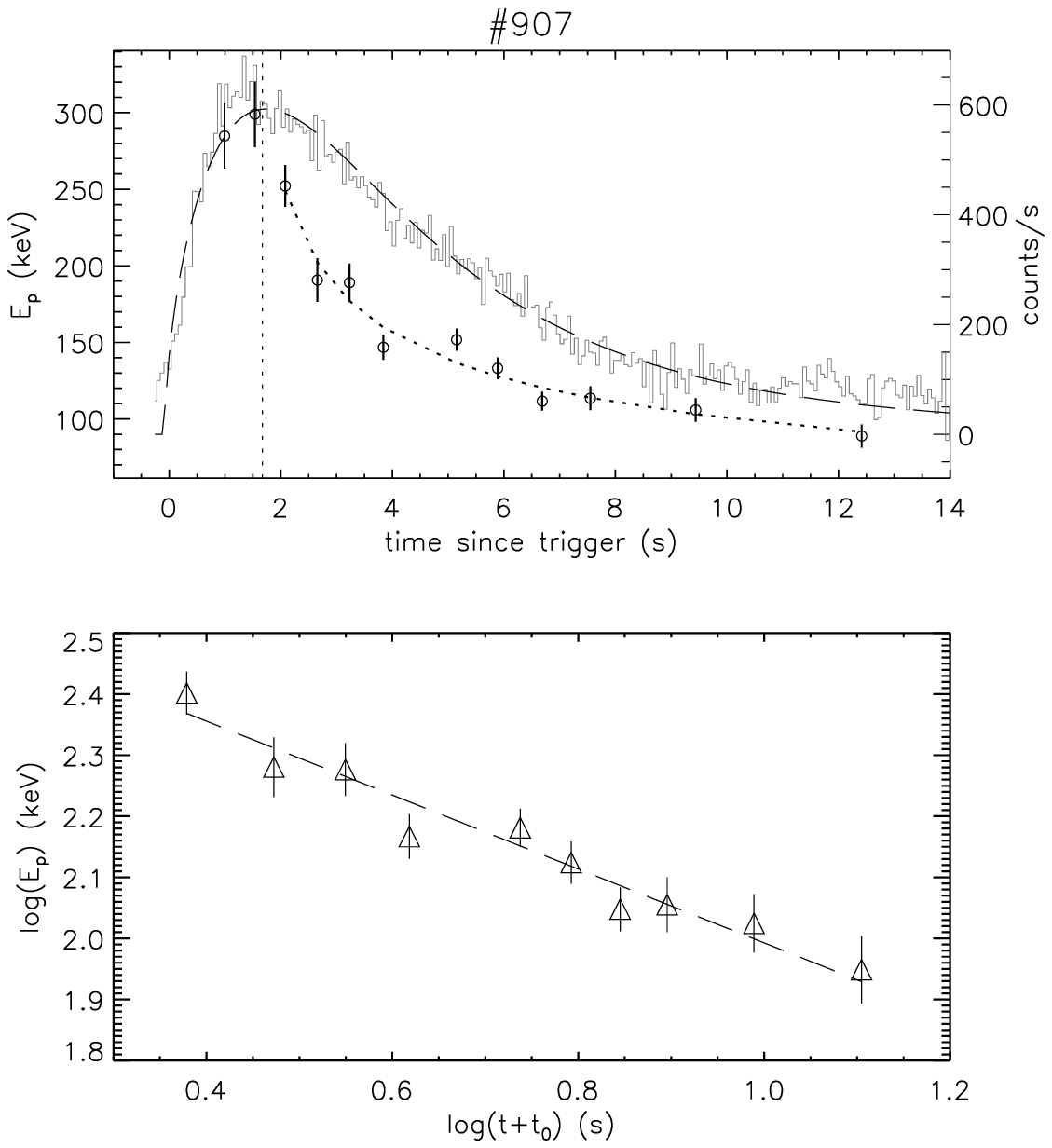}}
\resizebox{3in}{!}{\includegraphics{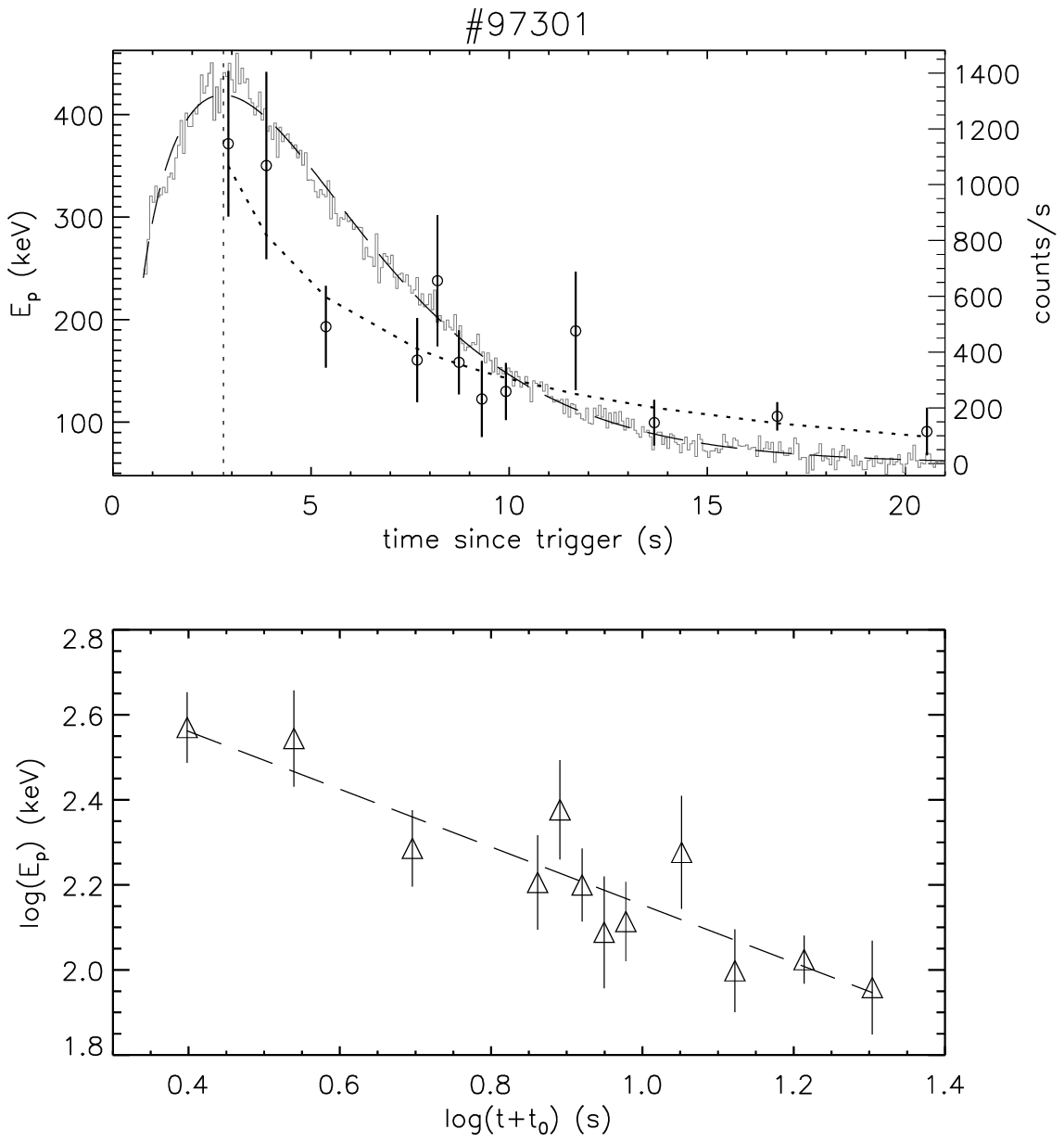}}
\resizebox{3in}{!}{\includegraphics{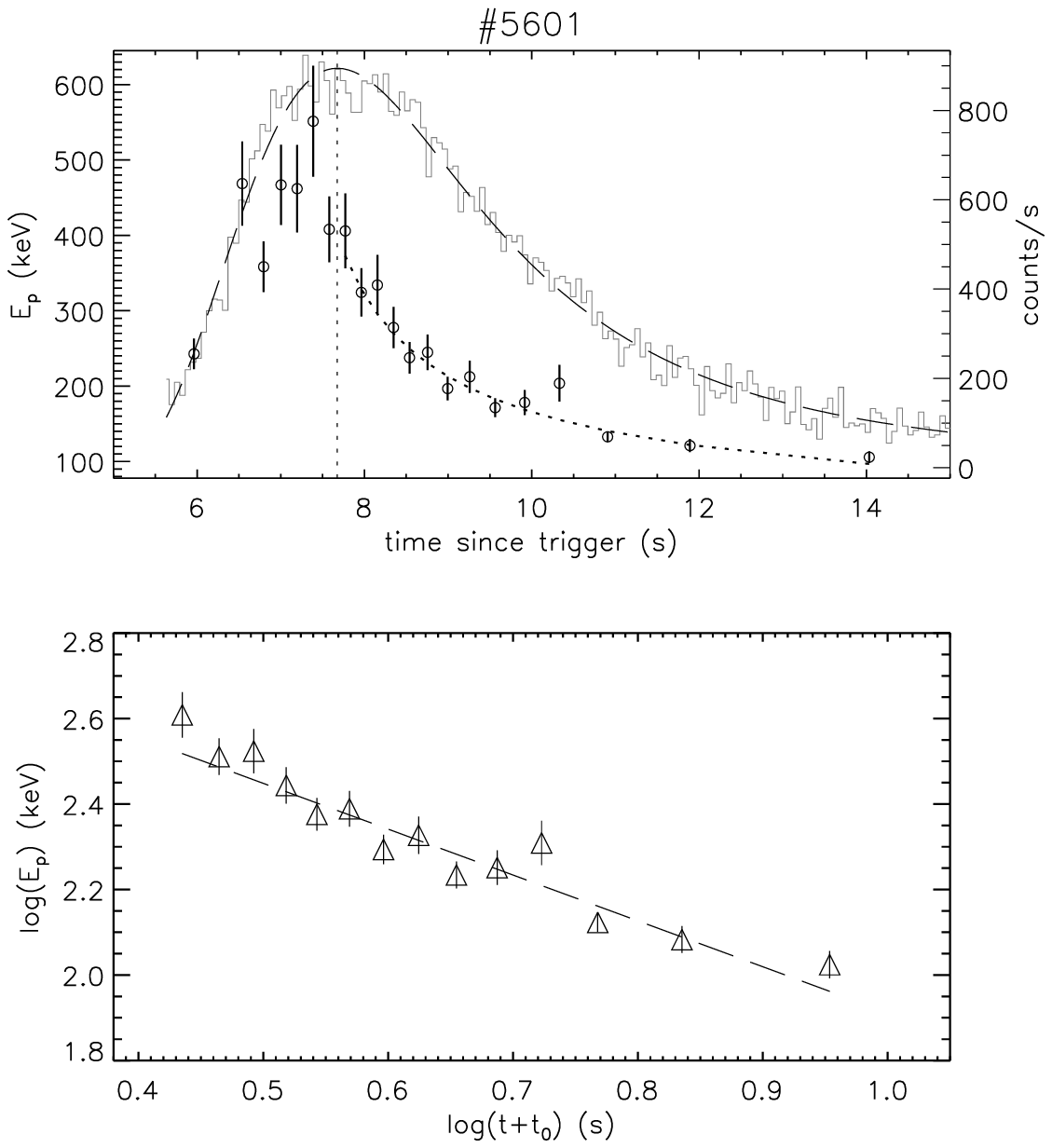}}
\resizebox{3in}{!}{\includegraphics{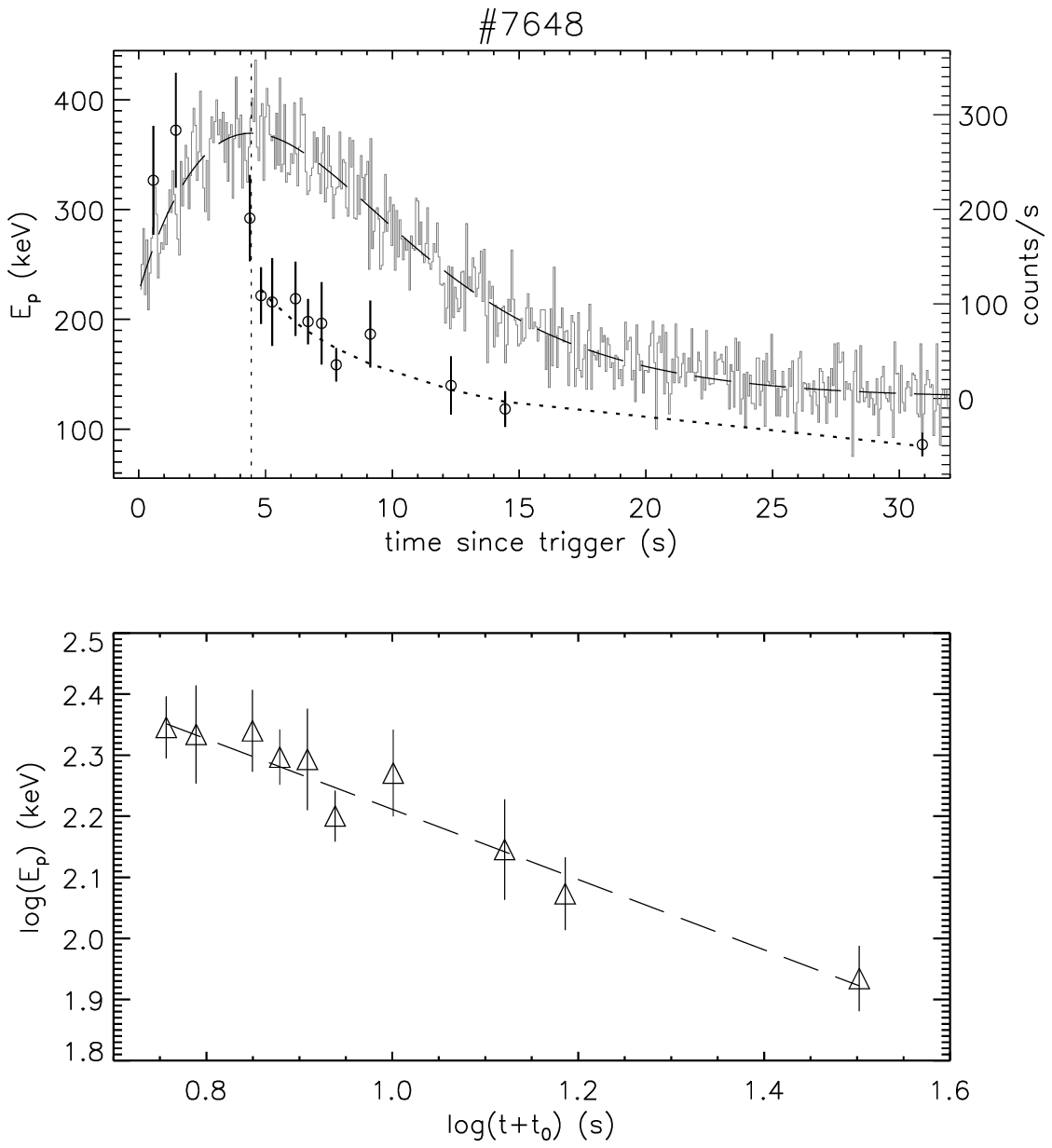}}
\caption{The example plots of spectral and temporal behaviors of the
pulses in sample 1. The top panels show the evolutions of the peak
energy $E_{p}$ of the time-resolved spectra (left axes) and their
corresponding pulse light curves (all four energy channels are used)
(right axes). The best fit of pulse is indicated with a long dashed
curve, the vertical dotted line represents the peak time of pulses,
while the dotted curve is the fitting line of the $E_{p}$ using a
single power-law function. The bottom panel indicates the
developments of the $E_{p}$ within the decay phase of the pulses and
the long dashed lines represent the best linear fitting.}
 \label{}
\end{figure}

\begin{figure}
\centering \resizebox{3in}{!}{\includegraphics{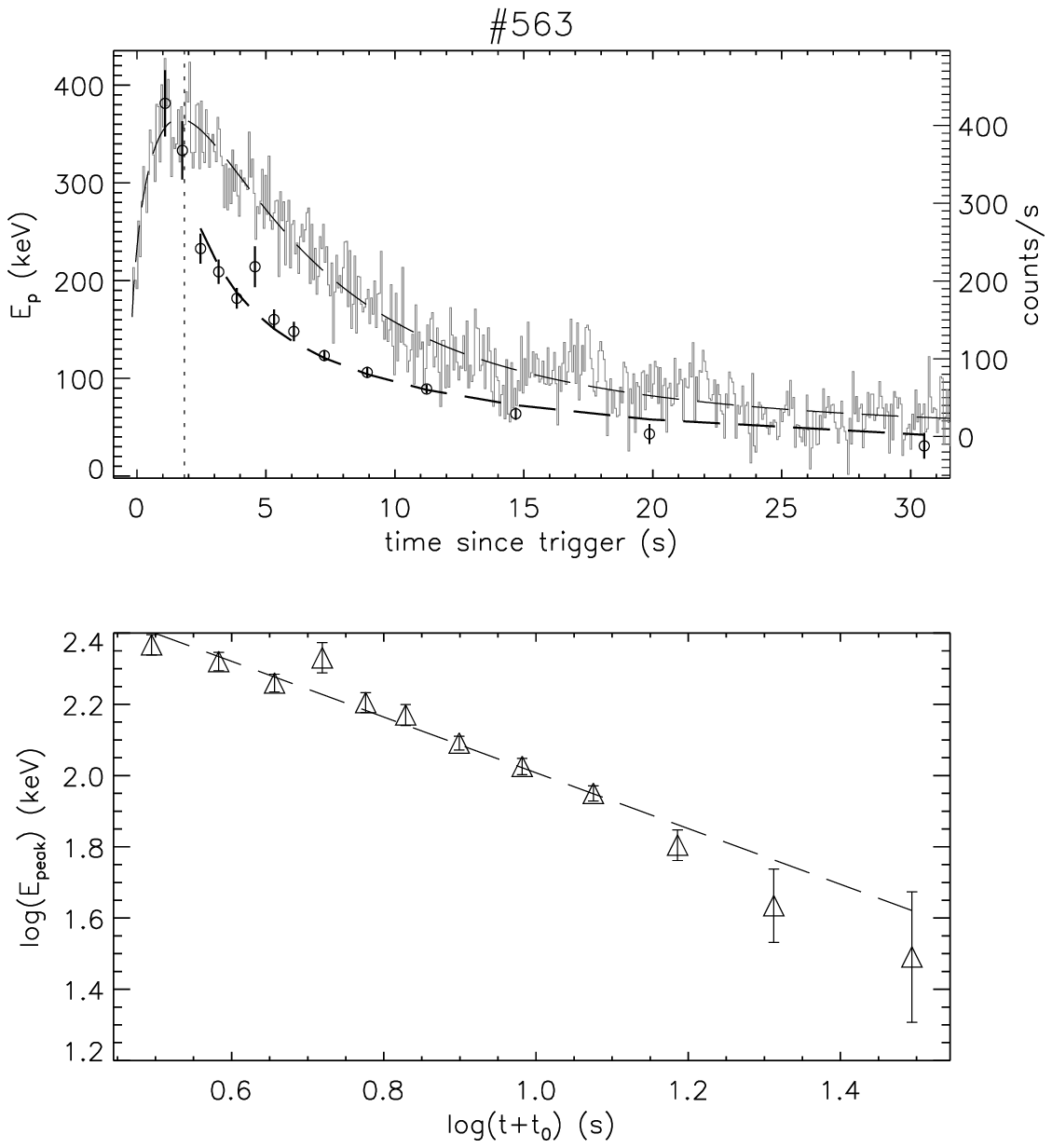}}
\resizebox{3in}{!}{\includegraphics{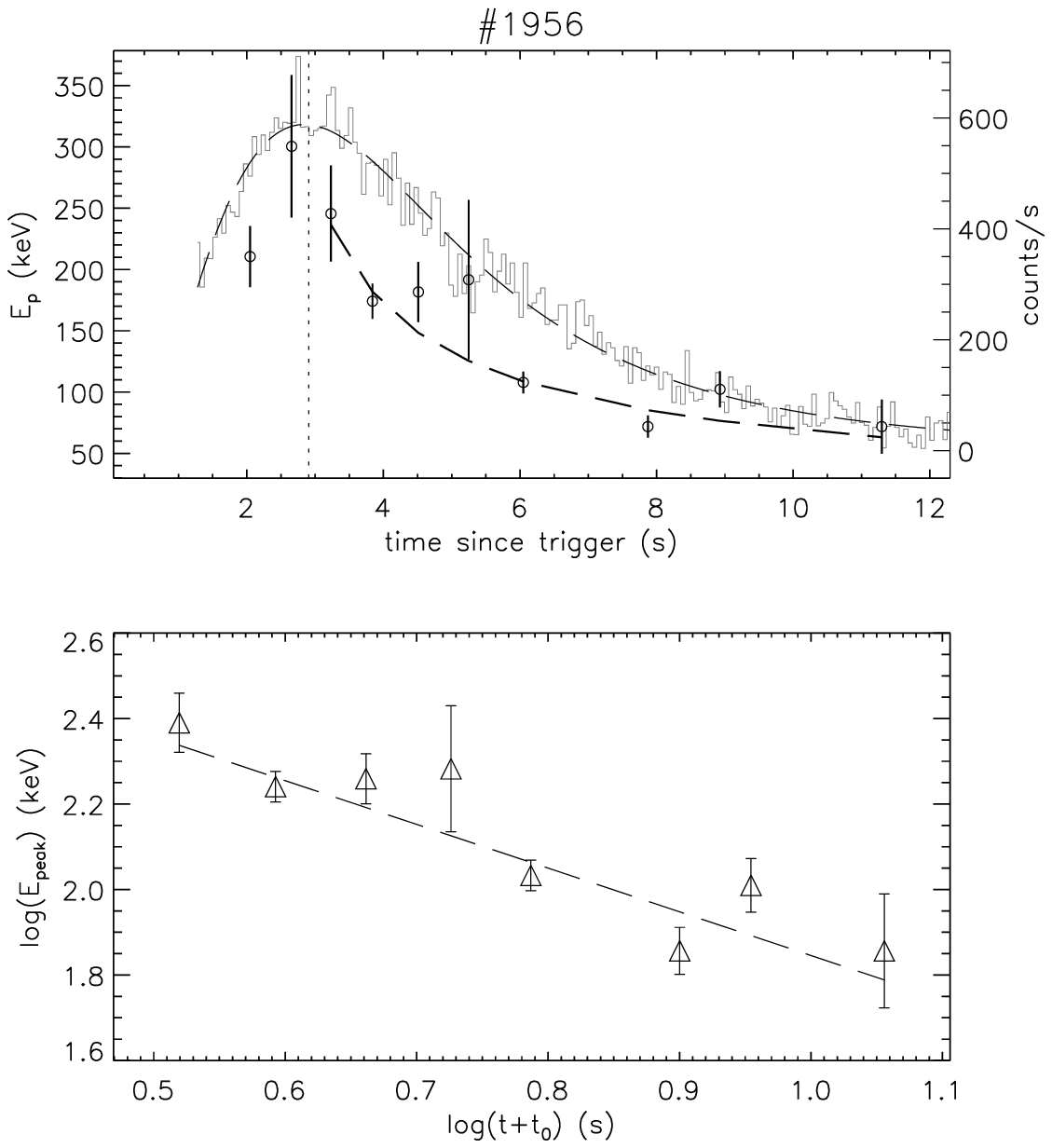}}
\resizebox{3in}{!}{\includegraphics{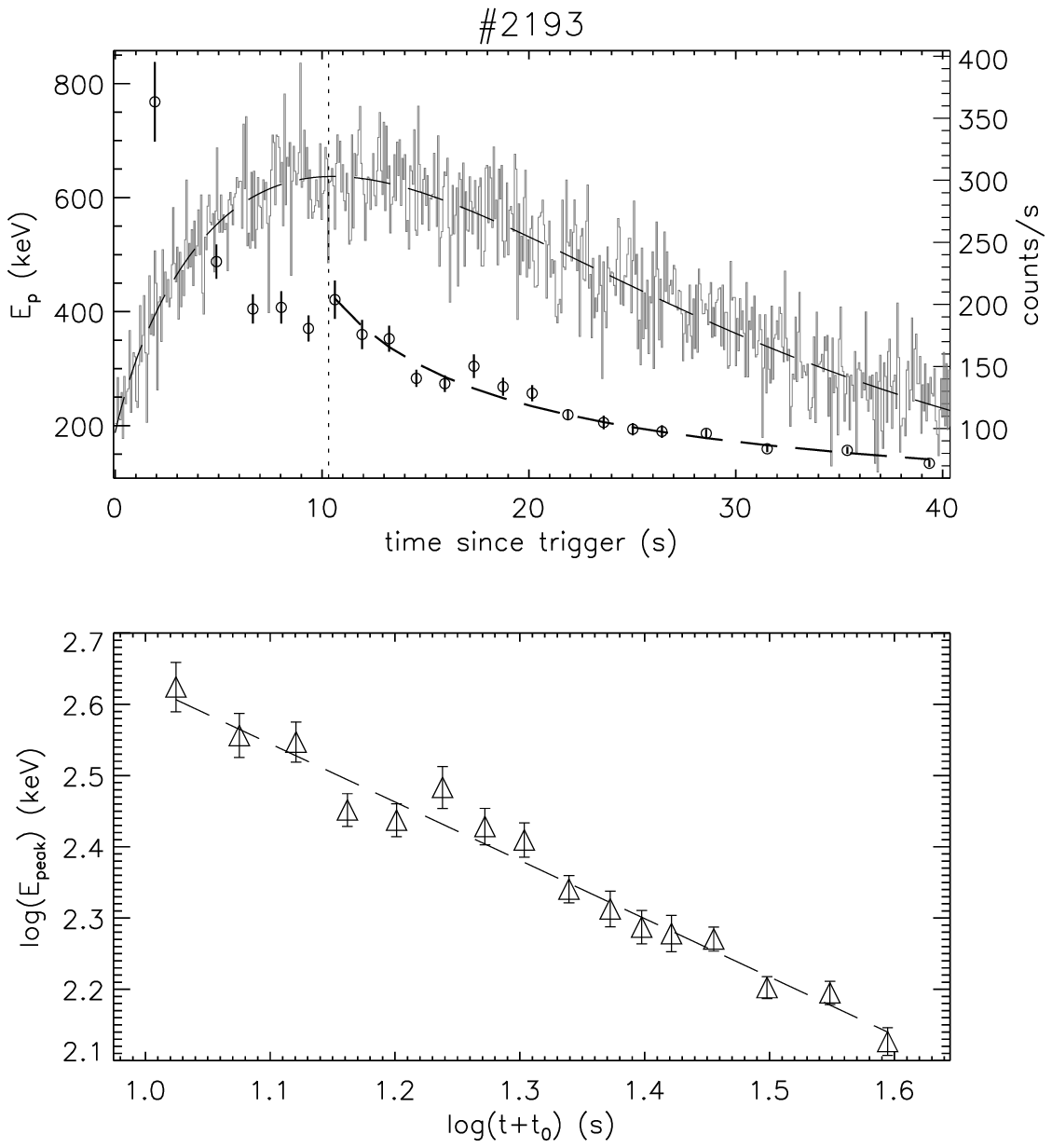}}
\resizebox{3in}{!}{\includegraphics{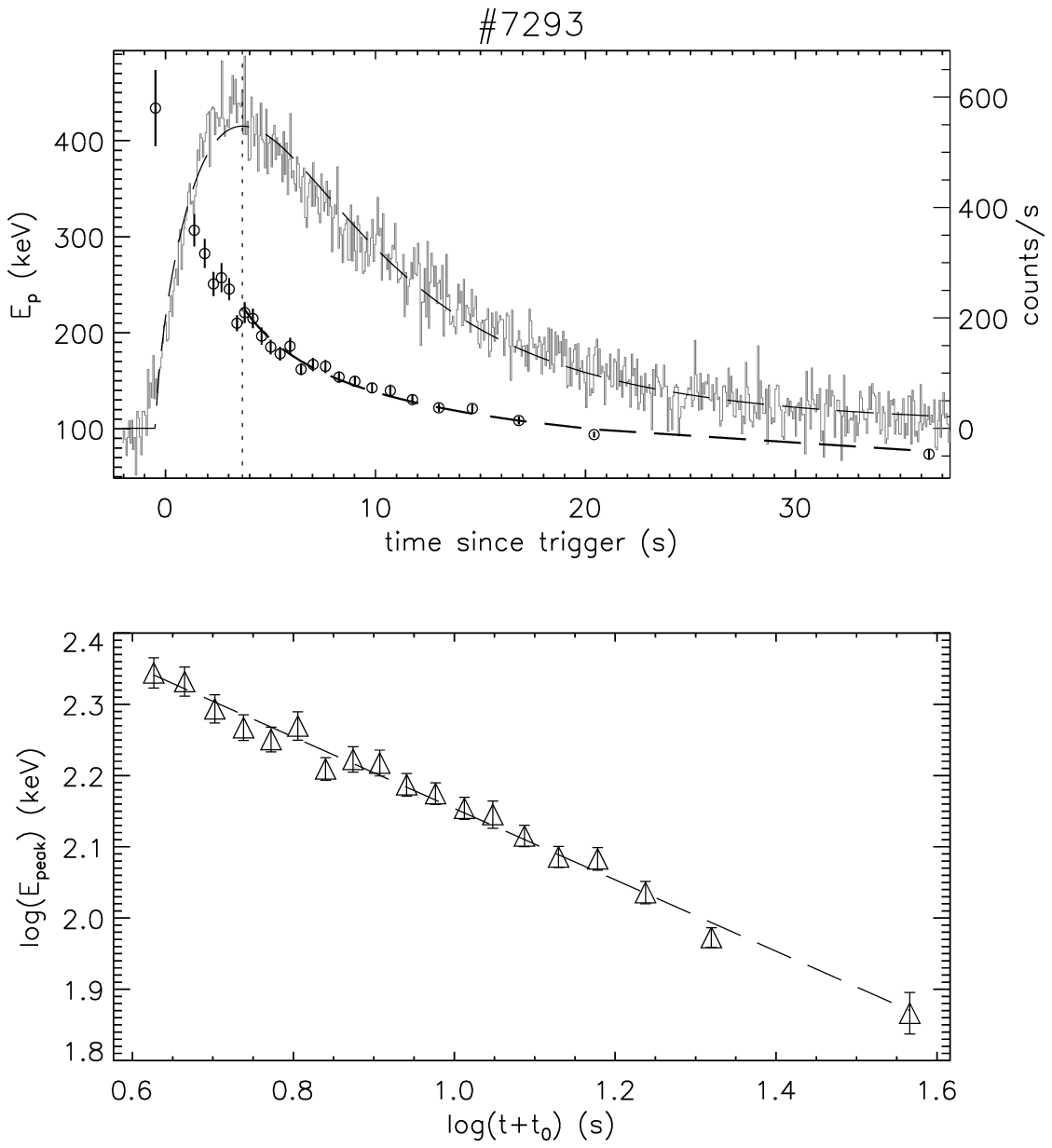}}
\caption{The example plots of spectral and temporal behaviors of the
GRB pulses in sample 2. The meaning of the symbols are the same as
figure 1. }
 \label{}
\end{figure}

\begin{figure}
\centering
\resizebox{3.2in}{!}{\includegraphics{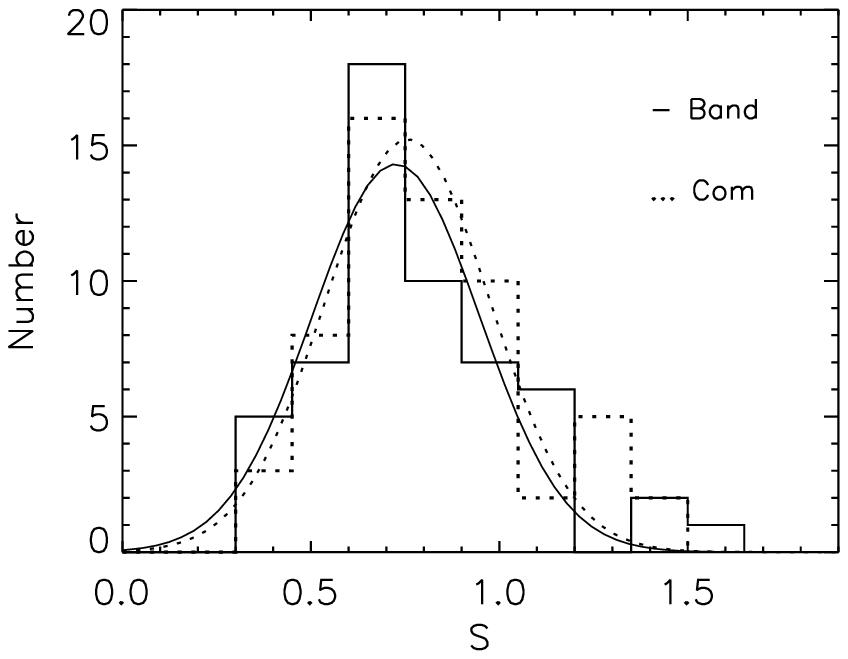}}
\resizebox{3.2in}{!}{\includegraphics{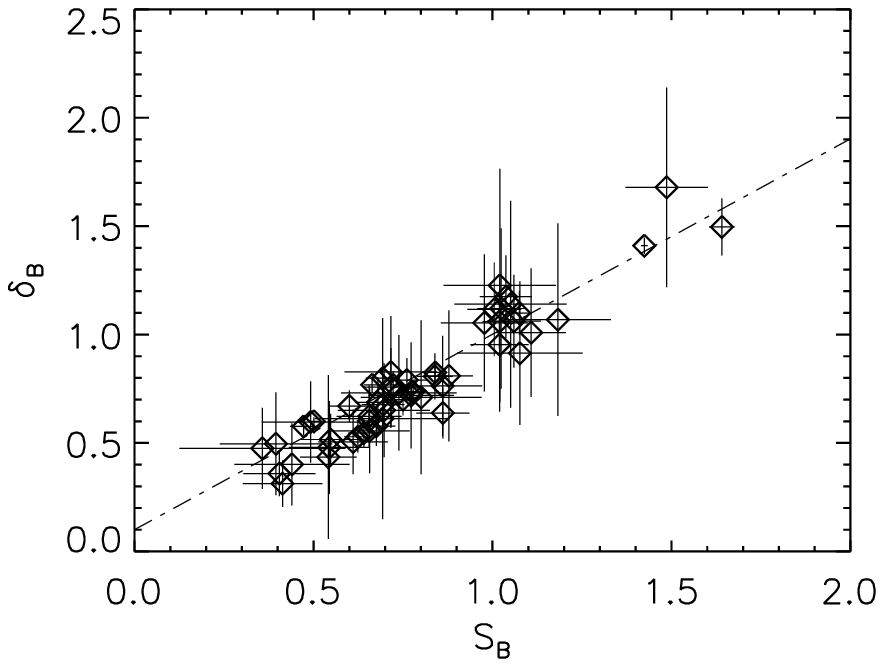}} \caption{The
distribution histograms of slope, S, (left panel) for sample 1
(solid line) and sample 2 (dashed line), where the curves represent
the Gaussian fit to the two distributions. The right panel is the
scatter plot of S vs. power-law index $\delta$ for sample 1.}
 \label{}
\end{figure}

In order to compare these results with the theoretical ones
presented by Paper I, we also plot them in Figures 1 and 2 in a
logarithmic manner for samples 1 and 2, respectively. In addition,
the evolutionary slope of the fitted peak energy $E_{p}$ should be
$d \log E_{p} /d \log(t+ t_{0})$ rather than $d \log E_{p} /d
\log(t+t_{trigger})$, where $t_{0}$ is the starting point of the
rising segment of the FRED pulse. For each pulse of our sample we
use LINFIT (Press 1992) to perform a linear least-squares fit at a 1
$\sigma$ confidence level to the two quantities and have $\log E_{p}
=W - S \log (t+t_{0})$. The evolutionary slope, S,  are listed in
Tables 1 and 2. We first check the consistence between the power-law
index $\delta$ and the slope S. Figure 3 (right panel) gives the
scatter plot of them, which shows they are indeed consistent.

The distribution histogram of S is indicated in Figure 3. We find
from Figure 3 for the two samples that: (i) the slopes vary from
0.23 to 1.66 for sample 1 and from 0.35 to 1.47 for sample 2; (ii)
the distributions of slopes are normal and the corresponding mean
values and standard deviations are around 0.73 and 0.22 for sample 1
and 0.76 and 0.22 for sample 2, respectively; (iii) the
corresponding values of the medians are 0.74 and 0.76 for samples 1
and 2, respectively. These indicate that for most of the pulses the
slopes are good consistent with that predicted by paper I, which
showed that the evolutionary slopes of the pulse decay phase are
$S_{1}= 0.95 \pm 0.01$ and $S_{2}= 0.75 \pm 0.01$ corresponding to
the intrinsic Band function spectrum and Comptonized or thermal
synchrotron spectrum, respectively. Therefore, within the limit of
uncertainty the corresponding intrinsic spectra of most of bursts
(about 25 for sample 1 and 29 for sample 2) may bear the intrinsic
Comptonized or thermal synchrotron spectrum, whereas only a small
fraction of bursts (about 16 for sample 1 and 12 for sample 2) may
be associated with the intrinsic Band spectra. There are also some
pulses (about 15 for sample 1 and 18 for sample 2) whose slopes are
inconsistent with both theoretical values, which is similar to the
result investigated by Paper I.

\subsection{the relation between evolutionary slope and the spectral parameters}

We suspect there is a relation between the first $E_{p}$ (we denote
it as $E_{p,max}$) of the decay pulse and the slope because we find
that: (i) the first $E_{p}$ is usually the maximal value of the
decay pulse; (ii) the larger the first $E_{p}$ is, the steeper of
the fitting curves is. Figure 4 demonstrates the relation between
$E_{p,max}$ and S for the the samples 1 (panel (a)) and 2 (panel
(b)). An correlation between the two quantities is identified for
both samples. For the sample 1 a liner regression analysis ($\log S=
A + B \times \log E_{p}$) yields the intercept $A=-1.78 \pm 0.08$
and the slope $B=0.73 \pm 0.03$, with the correlation coefficient
$R=-0.67$ $(N=56, p<10^{-4})$. While for the sample 2, the analysis
produces the intercept $A=-1.66 \pm 0.06$ and the slope $B=0.66 \pm
0.02$, with the correlation coefficient $R=-0.68$ ($N=59,
p<10^{-4}$).

\begin{figure}
\centering \resizebox{3.2in}{!}{\includegraphics{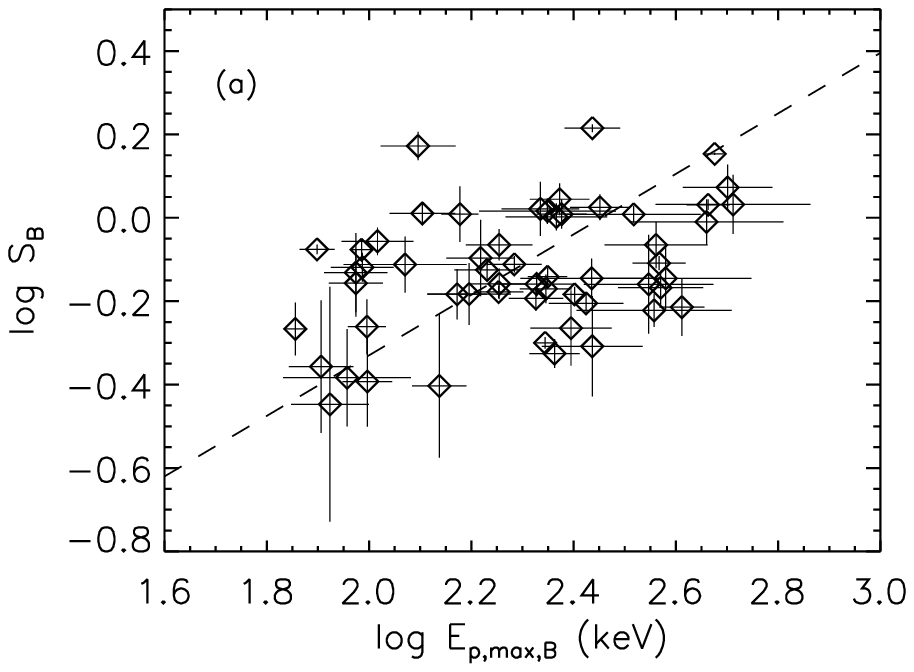}}
\resizebox{3.2in}{!}{\includegraphics{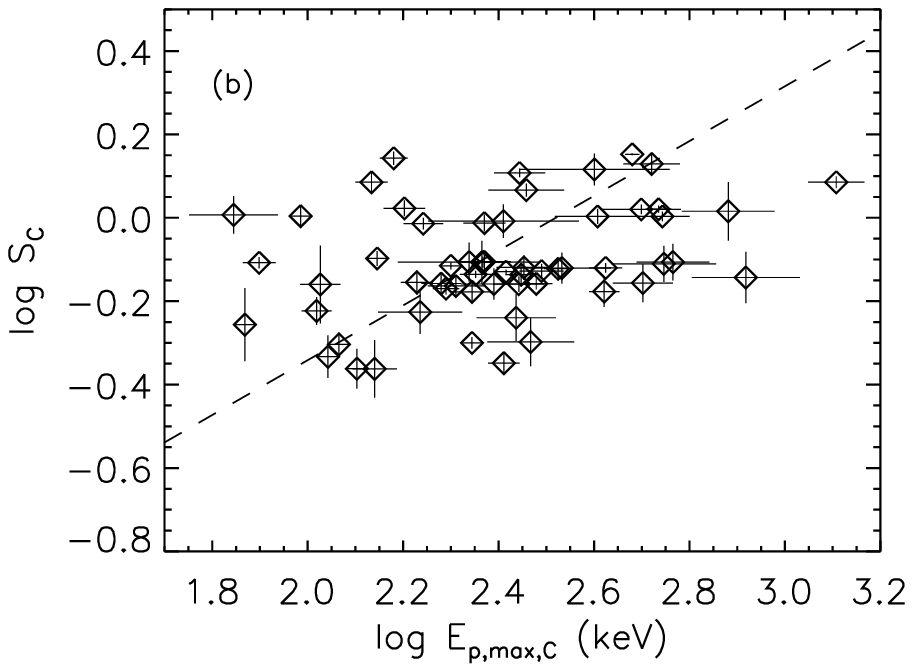}} \caption{The
evolutionary slope S vs. the $E_{p,max}$ of decay pulse for sample 1
(a) and sample 2 (b), where the long dashed lines represent the best
fitting lines.} \label{}
\end{figure}

Sakamoto et al. (2008) have showed that the $E_{p}$ is correlated
with the photon index ($\Gamma$) derived from a simple power-law
model. Motivated by this, we also examine if there are also
correlation between the $E_{p}$ and the S since the S corresponds to
the power-law index $\delta$. Figure 5 displays the $E_{p}$ of
time-integrated spectra versus S. A clear correlated relationship
between the two quantities is found. A straight line is fitted to
the points: (i) $\log S= -1.16(\pm 0.032)-0.50(\pm 0.01)\log E_{p}$,
with the correlation coefficient $R=0.65$ (N=56, $p<10^{-4}$) for
sample 1; (ii) $S=-1.22(\pm 0.022) - 0.51(\pm 0.01)E_{p}$, with the
correlation coefficient $R = 0.67$ (N= 59, $p<10^{-4}$) for sample
2.
\begin{figure}
\centering \resizebox{3.2in}{!}{\includegraphics{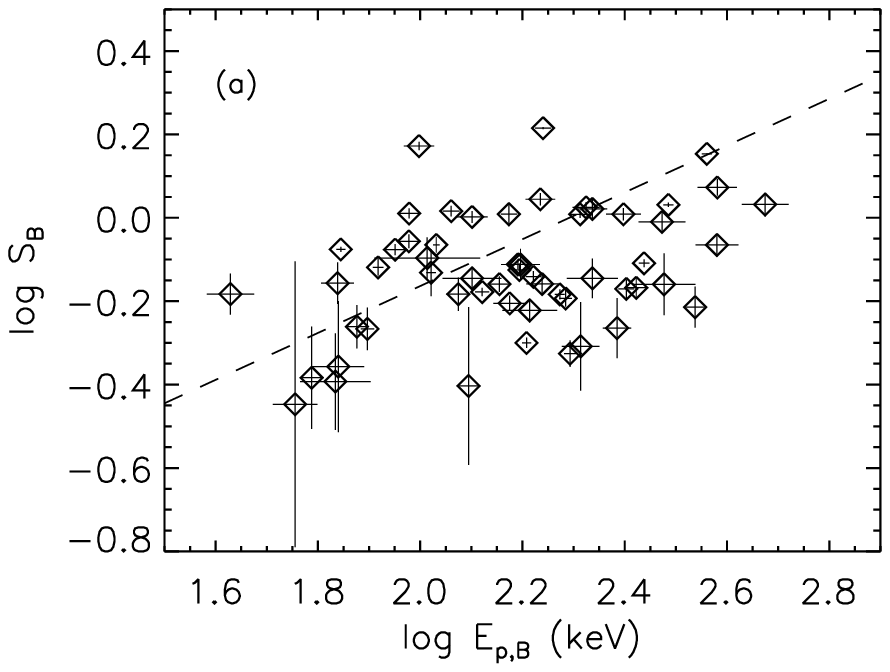}}
\resizebox{3.2in}{!}{\includegraphics{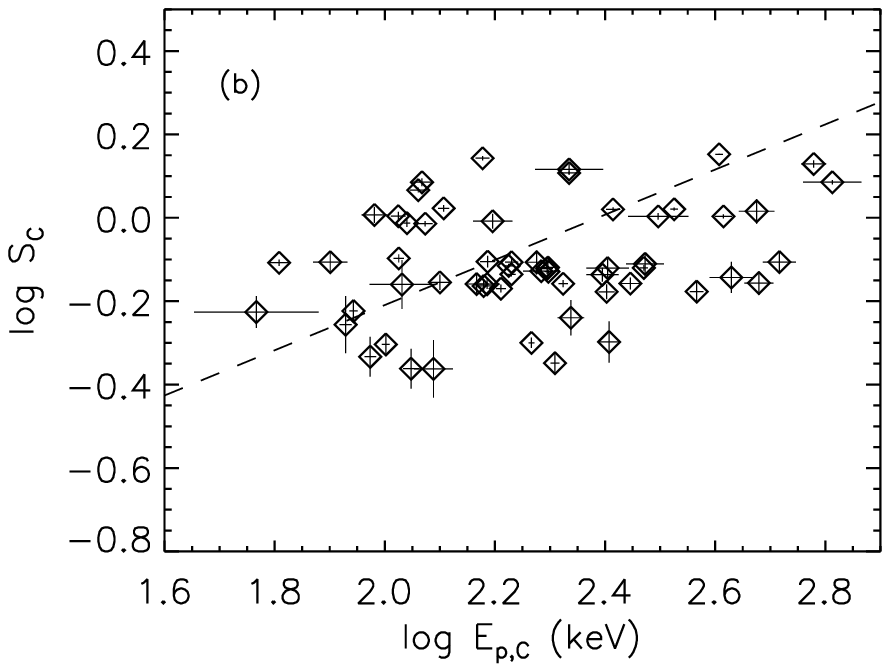}} \caption{The $\log$
S vs. $\log$$E_{p}$ for sample 1 (a) and sample 2 (b).}
 \label{}%
\end{figure}

From the Figures 4 and 5 we find that both the $E_{p, max}$ and the
$E_{p}$ of time-integrated spectra are correlated with the S. We
deduce that there must be some relation between them. Figure 6 gives
the scatter plot of the $E_{p,max}$ versus the $E_{p}$ of
time-integrated spectra, the corresponding correlation coefficient
$R = 0.94$ (N= 56, $p<10^{-4}$) for the sample 1 and $R = 0.92$ (N=
59, $p<10^{-4}$) for sample 2, which indicates there are indeed
strong correlation between them.

\begin{figure}
\centering \resizebox{3.2in}{!}{\includegraphics{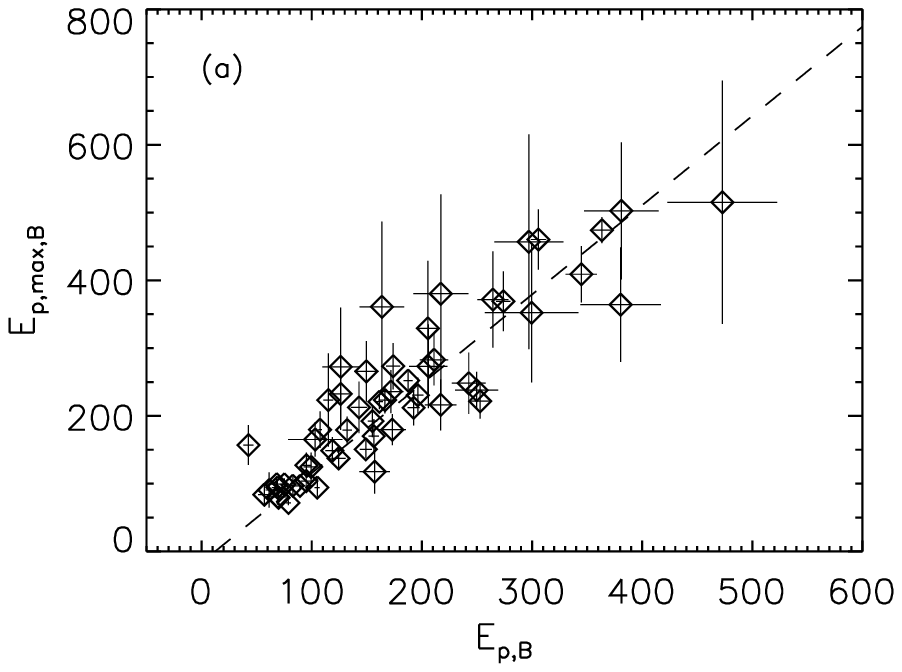}}
\resizebox{3.2in}{!}{\includegraphics{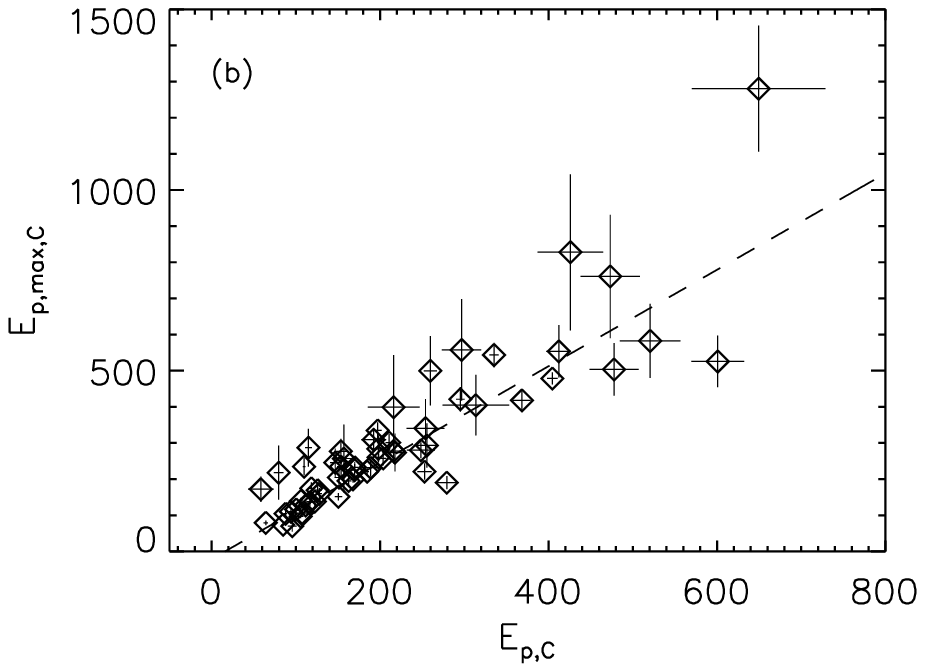}} \caption{$E_{p}$ of
time-integrated spectra vs. the $E_{p,max}$  for sample 1 (a) and
sample 2 (b). The strong correlation is indicated for both samples.}
\label{}%
\end{figure}

We also examine the relation between the S and lower energy index
$\alpha$ as well as photon flux, $f$, (the energy range is 25-1800
keV). Figure 7 shows the plot of the $\alpha$ versus S. A
anticorrelation is suggested for both of samples with the
correlation coefficient $R = -0.57$ (N= 56, $p<10^{-4}$) for the
sample 1 and $R = -0.44$ (N= 59, $p=4.15\times10^{-4}$) for sample
2.

\begin{figure}
\centering \resizebox{3.2in}{!}{\includegraphics{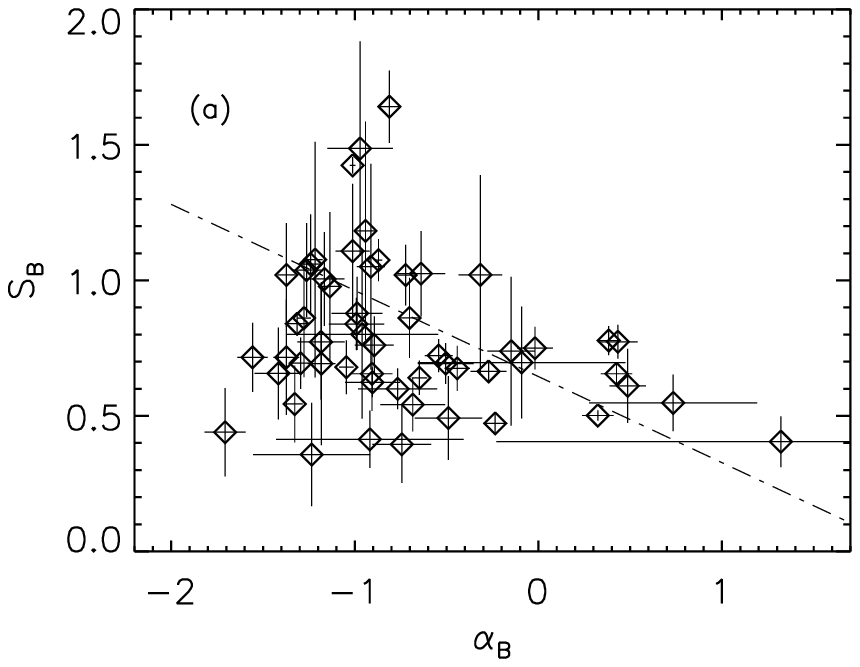}}
\resizebox{3.2in}{!}{\includegraphics{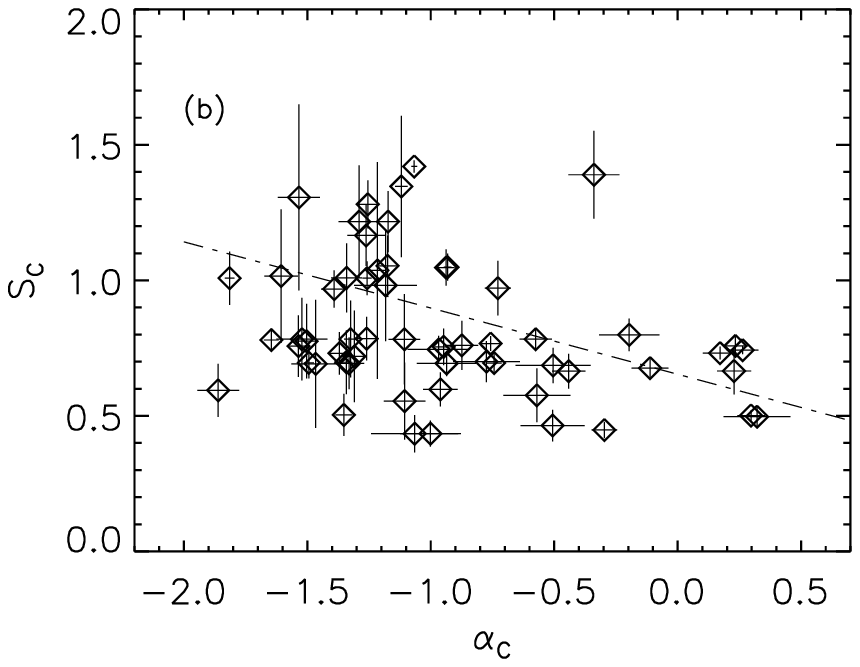}}
\caption{Evolutionary slope S vs. low energy index $\alpha$ for
sample 1 (a) and sample 2 (b), where the long dashed lines represent
the best fitting lines.}
 \label{}%
\end{figure}

\begin{figure}
\centering \resizebox{3.2in}{!}{\includegraphics{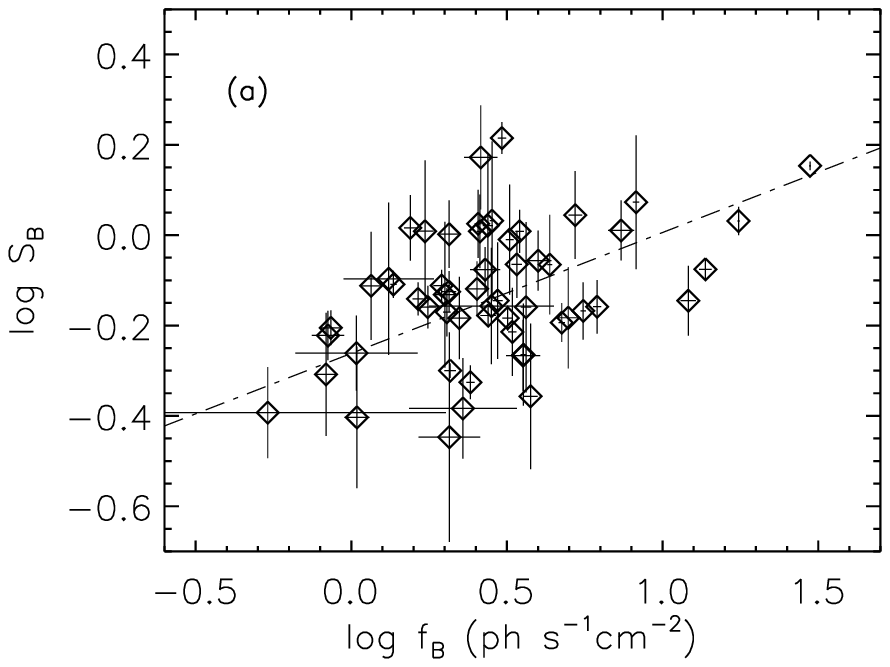}}
\resizebox{3.2in}{!}{\includegraphics{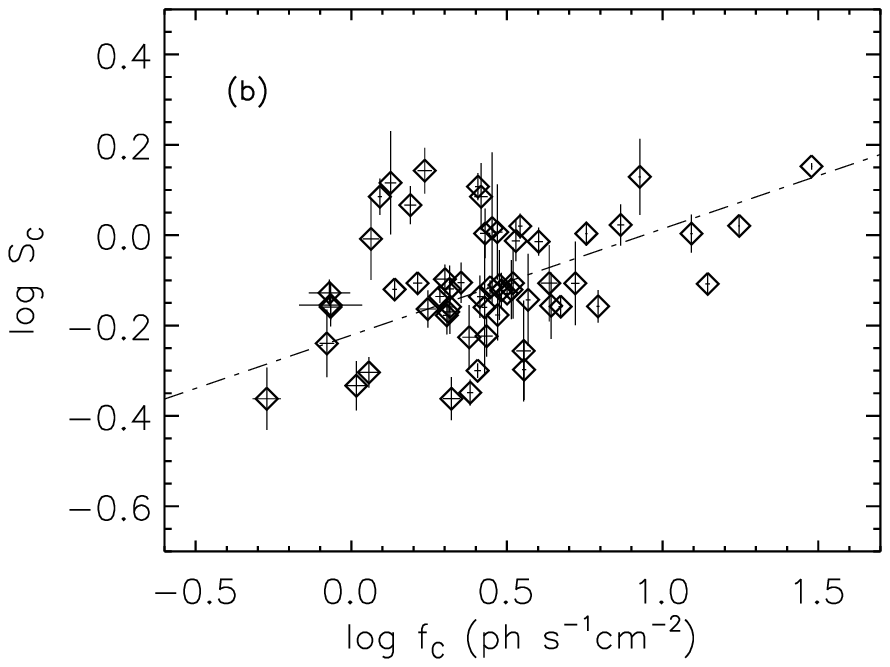}}
\caption{Evolutionary slope S vs. photon flux $f$ for sample 1 (a)
and sample 2 (b), where the long dashed lines represent the best
fitting lines. }
 \label{}%
\end{figure}

Figure 8 plots the $f$ versus S. It is found that S is also
correlated with $f$. The regression analysis give the best fitting
line: (i) $\log S= -0.26 (\pm 0.01)+0.27 (\pm 0.01)\log f$, with the
correlation coefficient $R=0.82$ (N=56, $p<10^{-4}$) for sample 1;
(ii) $\log S=-0.22(\pm 0.02) + 0.24(\pm 0.01)\log f$, with the
correlation coefficient $R = 0.80$ (N= 59, $p<10^{-4}$) for sample
2.



Since the S is related to the $E_{p}$, we wonder if there are also
correlation between the S and intrinsic peak energy $E_{p,i}$. To
verify the relation we need to estimate the redshifts of our
selected bursts due to unknown redshift information of BATSE bursts.
We attempt to use the pseudo-redshift derived by Yonetoku et al.
(2004). This pseudo-redshifts were estimated based on a new relation
between the spectral peak energy $E_{p}$ and the 1 s peak
luminosity, which was derived by combining the data of $E_{p}$ and
the peak luminosities by BeppoSAX and BATSE. In addition, it looks
considerably tighter and more reliable than the relations suggested
by the previous works (e.g. Amati et al. 2002). The intrinsic peak
energy, $E_{p,i}$, in the rest frame is related with the observed
$E_{p}$ in the observer frame by $E_{p,i} = E_{p}(1 + z)$. There are
49 and 53 pulses with pseudo-redshift information in samples 1 and
2, respectively. The estimated relation between them is displayed in
Figure 9.

\begin{figure}
\centering \resizebox{3in}{!}{\includegraphics{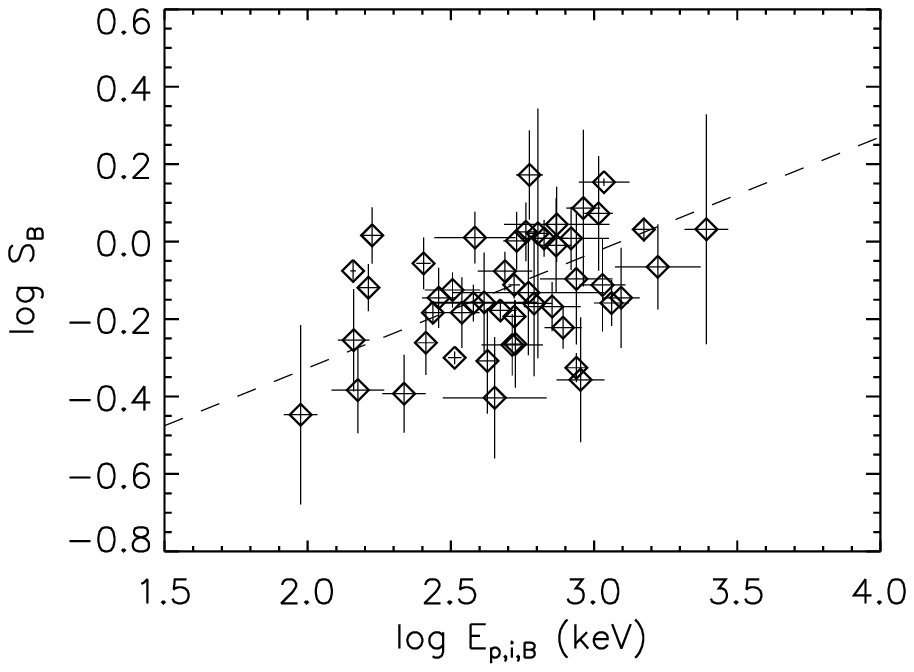}}
\resizebox{3in}{!}{\includegraphics{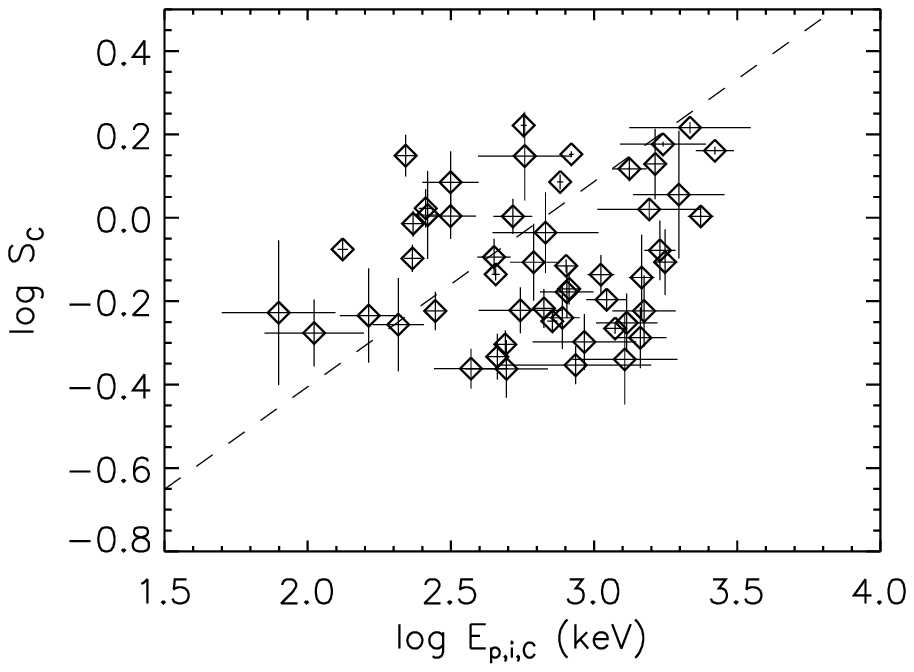}} \caption{The
evolutionary slope S vs. intrinsic peak energy $E_{p,i}$ for sample
1 (a) and sample 2 (b), where the long dashed lines represent the
best fitting lines.} \label{}
\end{figure}

An correlated relationship between the two quantities is also
identified for both samples. The best fit functional form of this
relation is $\log S=-0.923(\pm 0.047) -0.298(\pm 0.017)\log
E_{p,i}$, with the correlation coefficient $R=0.67$ (N=50,
$p<10^{-4}$) for sample 1. For the sample 2 $\log S=-1.388(\pm
0.061)- 0.492(\pm 0.022)\log E_{p,i}$, with the correlation
coefficient $R = 0.50$ (N= 53, $p<10^{-4}$). Comparing it with the
case of observed relation between S and $E_{p}$ the intrinsic
relation seems more loosely defined for sample 2.

\begin{deluxetable}{ccccccccccccc}   %
\tabletypesize{\scriptsize}  \tablecaption{A list of burst sample
and various parameters for sample 1.
\label{tbl-1}}\tablewidth{0pt}\rotate \tablehead{ \colhead{Trigger}
& \colhead{$F_{p}$} &\colhead{time interval} & \colhead{n}&
\colhead{$t_{0}$} & \colhead{$t_{m}$} & \colhead{$\chi_{\nu}^{2}1$}
&
\colhead{$E_{p}$}& \colhead{$\alpha$}& \colhead{$\delta$} & \colhead{S} & \colhead{$\chi_{\nu}^{2}2$}\\
&(phs cm$^{-2}$s$^{-1}$)& (s) & & (s) &(s)& & (keV)& & }\startdata
&  &  &&  & Band photon model &  & & & &  \\
\hline \hline
\\
563 & 1.89 $\pm$ 0.14 & 1.84-15.00 & 9  &  0.66 $\pm$ 0.04  & 1.84 $\pm$ 0.05 & 1.25 & 166.53 $\pm$ 5.74 & -0.54 $\pm$ 0.07& 0.77 $\pm$ 0.28 & 0.72 $\pm$ 0.09 & 1.34 \\
907 &  3.57 $\pm$ 0.17 & 1.67-15.00 & 10 &  0.31 $\pm$ 0.04  & 1.67 $\pm$ 0.03 & 1.38 & 187.34 $\pm$ 4.10& 0.43 $\pm$ 0.08& 0.57 $\pm$ 0.07  &0.66 $\pm$ 0.06 & 1.77 \\
914 & 2.53  $\pm$ 0.16 &0.61-5.00 & 5 & 1.17 $\pm$ 0.45 & 0.61 $\pm$ 0.02  & 0.92&99.46 $\pm$ 6.77&-0.97 $\pm$ 0.18 & 1.53 $\pm$ 0.38  & 1.48 $\pm$ 0.40 & 0.15 \\
973$_{-}$1 & 5.29 $\pm$ 0.20 & 2.79-22.00 & 12 & -0.41 $\pm$ 0.07  & 2.79 $\pm$ 0.03 &1.34& 264.52 $\pm$ 14.11& -1.05 $\pm$ 0.04 & 0.68 $\pm$ 0.09 & 0.69  $\pm$ 0.12 & 0.56 \\
973$_{-}$2 & 5.29 $\pm$ 0.20 & 24.09-35.00 & 6 & -22.14 $\pm$ 0.01 & 24.09 $\pm$ 0.03 & 1.24& 217.21 $\pm$ 24.92 & -1.37  $\pm$ 0.07 & 0.83 $\pm$ 0.24 & 0.72 $\pm$ 0.21 & 0.25\\
999 & 11.55 $\pm$ 0.31 & 3.99-6.00 & 9 & -3.22 $\pm$ 0.14 & 3.99 $\pm$ 0.00 & 2.37&381.13 $\pm$ 33.72& -0.94 $\pm$ 0.05 & 1.05 $\pm$ 0.33 & 1.18   $\pm$ 0.41 & 0.51 \\
1406 & 1.97 $\pm$ 0.13  & 3.35-20.00 & 10 & 0.95 $\pm$ 0.07 & 3.35 $\pm$ 0.05 & 1.07 &118.71 $\pm$ 6.31 &  -0.91 $\pm$ 0.11 & 0.65 $\pm$ 0.13 & 0.62 $\pm$ 0.13 & 0.63 \\
1883 & 5.20 $\pm$ 0.18 &1.31-7.00 & 6 & 0.25 $\pm$ 0.07 & 1.31 $\pm$ 0.02 & 0.97& 249.72 $\pm$ 19.61 &-1.372 $\pm$ 0.05 & 0.94 $\pm$ 0.27 &1.01 $\pm$ 0.20 &  0.56 \\
1956 & 2.57 $\pm$ 0.13 & 2.91-15.00 &  8 &  0.07 $\pm$ 0.29& 2.91 $\pm$ 0.04 & 1.21 &126.23 $\pm$ 8.82 & -1.17 $\pm$  0.11& 1.11 $\pm$ 0.14 & 1.00 $\pm$ 0.17 & 0.95\\
1989 & 2.73 $\pm$ 0.15 & 116.54-130.00 & 5 & -111.58 $\pm$ 0.20 & 116.54 $\pm$ 0.05 & 1.32&89.32 $\pm$ 4.93&  -0.99 $\pm$ 0.15 & 0.80 $\pm$ 0.05  & 0.84 $\pm$ 0.09 & 1.11\\
2083$_{-}$1 & 45.42  $\pm$ 0.46 & 1.05-6.00 & 17  & 0.36 $\pm$ 0.22& 1.04 $\pm$ 0.00& 5.98&363.25  $\pm$  8.24  &-1.40 $\pm$ 0.02 & 1.24 $\pm$ 0.09 & 1.42 $\pm$ 0.03 & 1.15\\
2083$_{-}$2 & 45.42  $\pm$0.46 & 8.68-20.00 & 42 & -4.02 $\pm$ 0.78 & 8.68 $\pm$ 0.01 & 2.26&69.95 $\pm$ 1.40&-1.31 $\pm$ 0.07 & 0.83 $\pm$ 0.34 & 0.84 $\pm$ 0.04 & 0.79 \\
2138 & 7.00  $\pm$ 0.20 & 1.29-15.00 & 11& 1.28 $\pm$ 0.21 & 1.29 $\pm$ 0.06 &1.17& 205.92  $\pm$ 17.4&  -0.49  $\pm$ 0.18 & 0.59 $\pm$ 0.02 & 0.49 $\pm$ 0.15 & 0.58\\
2193 & 1.55  $\pm$ 0.13 &10.32-42.00 & 15 &-0.04 $\pm$ 0.5 & 10.32 $\pm$ 0.24 & 0.98 &274.11 $\pm$ 6.48 & 0.38  $\pm$ 0.06& 0.74 $\pm$ 0.46 & 0.78 $\pm$ 0.05 & 1.08\\
2387 & 3.86  $\pm$ 0.16 & 6.49-40.00 & 38 & -0.53 $\pm$ 0.27 &6.49 $\pm$ 0.07 &1.15&132.22 $\pm$ 4.17 &-0.27 $\pm$ 0.09 & 0.76 $\pm$ 0.04 & 0.66 $\pm$ 0.04 & 1.48\\
2484 & 1.55  $\pm$ 0.13 & 2.01-10.00 & 6 & 2.66 $\pm$ 0.85 & 2.01 $\pm$ 0.07 &  1.09&107.55 $\pm$ 1.60 & -0.70 $\pm$ 0.06& 0.67 $\pm$ 0.13 & 0.86 $\pm$ 0.15 & 0.85\\
2519 & 1.53  $\pm$ 0.14 & 0.59-15.00 & 13& 0.45 $\pm$ 0.02 &0.59 $\pm$ 0.06 & 1.02 &163.76 $\pm$ 20.23 &  -0.77 $\pm$  0.22& 0.67 $\pm$ 0.07 & 0.60 $\pm$ 0.07 & 1.00\\
2662 & 1.52  $\pm$ 0.14 & 1.31-12.00 & 6 & 0.30 $\pm$ 0.05 &1.31 $\pm$ 0.07 &1.11&149.64  $\pm$  11.03 & -0.90 $\pm$ 0.15 &0.53 $\pm$ 0.07 & 0.62 $\pm$ 0.09 & 1.15\\
2665 & 1.99  $\pm$ 0.15 & 1.42-8.00  & 7 & 1.22 $\pm$ 0.32  &1.42 $\pm$ 0.05 & 0.89&105.03 $\pm$  2.25 &  -0.15 $\pm$ 0.13 & 0.73 $\pm$ 0.22 & 0.74 $\pm$ 0.25 & 0.50\\
2700 & 4.06  $\pm$ 0.18 & 53.7-61.00 & 14 & -52.84 $\pm$ 0.01 & 53.71 $\pm$ 0.03 &1.37&242.36  $\pm$  15.13 &  -1.33 $\pm$  0.04 & 0.61 $\pm$ 0.21 & 0.54 $\pm$ 0.14 & 0.12\\
2880 & 2.90  $\pm$ 0.14 &0.47-3.00  & 6 & 0.34 $\pm$ 0.04 &0.47 $\pm$ 0.01&1.43&124.33 $\pm$ 6.46 & -0.74 $\pm$  0.16 &0.46 $\pm$ 0.12 & 0.50 $\pm$ 0.14 & 0.40\\
2919 & 5.77  $\pm$ 0.19 &0.34-7.00 & 9 & 1.57  $\pm$ 0.12 &0.34 $\pm$ 0.02 & 1.39&380.43 $\pm$ 36.64 & -1.28 $\pm$  0.04 & 0.79 $\pm$ 0.19 & 0.86 $\pm$ 0.29 & 0.82\\
3003 & 2.83 $\pm$ 0.16  & 9.68-24.00 & 10 & 8.94 $\pm$ 7.15 &9.68  $\pm$ 0.08 &1.02&472.83  $\pm$ 49.82 & -1.22  $\pm$ 0.04 & 0.93 $\pm$ 0.22 & 1.07 $\pm$ 0.73 & 0.40\\
3143 & 2.59  $\pm$ 0.14 & 0.69-4.00 & 8 & 0.64 $\pm$ 0.25 & 0.69 $\pm$ 0.02 & 0.96&99.09  $\pm$  16.27 & -0.85 $\pm$  0.39 & 0.77 $\pm$ 0.44 & 0.80 $\pm$ 0.39 & 1.43\\
3256 & 1.76  $\pm$ 0.11 & 1.38-9.00 & 20 & 0.55 $\pm$ 0.08 &1.38 $\pm$ 0.07& 1.06 &142.93   $\pm$  7.39 & -0.50 $\pm$ 0.15 & 0.65 $\pm$ 0.05 & 0.70 $\pm$ 0.07 & 0.90\\
3257 & 3.06  $\pm$ 0.13 & 3.51-35.00 & 27 & 0.60 $\pm$ 0.06 & 3.51 $\pm$ 0.06 & 1.10&196.33 $\pm$ 5.15 &  -0.24 $\pm$ 0.06& 0.45 $\pm$ 0.02 & 0.57 $\pm$ 0.04 & 1.50\\
3290  & 10.71  $\pm$ 0.18 & 2.98-4.00 & 5& -2.73 $\pm$ 0.00 &2.98 $\pm$ 0.01& 2.52& 50.84 $\pm$ 10.53 & -1.69 $\pm$ 0.21 & 0.41 $\pm$ 0.30 & 0.45 $\pm$ 0.16 & 0.60\\
3415 & 9.16  $\pm$ 0.19 & 11.56-15.00  & 5 & -10.84 $\pm$ 0.02 & 11.56 $\pm$ 0.01&1.44&245.74 $\pm$ 28.54 & -1.21  $\pm$ 0.07 & 0.61 $\pm$ 0.33 & 0.70 $\pm$ 0.30 & 1.83\\
3648$_{-}$2 & 5.70  $\pm$ 0.15 & 23.85-36.00 & 5 & -19.47 $\pm$ 0.19 & 23.85 $\pm$ 0.18& 0.97&75.22 $\pm$ 3.68&0.73 $\pm$ 0.45 & 0.52 $\pm$ 0.08 & 0.55 $\pm$ 0.10 & 1.50\\
3648$_{-}$3 & 5.70  $\pm$ 0.15  & 41.04-48.00 & 12 & -32.20 $\pm$ 1.44 & 41.04 $\pm$ 0.02 & 1.36&173.94 $\pm$ 6.30& -0.81 $\pm$ 0.05& 1.49 $\pm$ 0.02 & 1.62 $\pm$ 0.09 & 0.85\\
3765 & 25.29  $\pm$ 0.27 &66.15-73.00 & 12 &-63.70 $\pm$ 0.24 & 66.15 $\pm$ 0.01 & 1.45&294.73 $\pm$ 7.20 &  -1.09 $\pm$  0.02 & 1.03 $\pm$ 0.19 &1.07 $\pm$ 0.08 & 2.05\\
3870 & 13.90  $\pm$ 0.23 & 0.50-6.00 & 9  & 0.46 $\pm$ 0.38  & 0.50 $\pm$ 0.01& 1.32&118.03  $\pm$ 19.34 & -1.61 $\pm$ 0.09& 0.76 $\pm$ 0.06  & 0.72 $\pm$ 0.14 & 0.72\\
3875 &2.80  $\pm$ 0.11  & 0.27-3.00 & 8  & 0.14 $\pm$ 0.01  & 0.27 $\pm$ 0.02 &1.61&69.64 $\pm$  4.02 & -1.23 $\pm$ 0.25& 0.47 $\pm$ 0.21 & 0.36 $\pm$ 0.19 & 1.14\\
3954 &  8.19 $\pm$ 0.19  & 0.78-6.00 & 8 & 2.26 $\pm$ 1.47 & 0.78 $\pm$ 0.01 & 1.05&297.13  $\pm$31.34 & -1.14 $\pm$ 0.06& 1.05 $\pm$ 0.08 & 0.98 $\pm$ 0.27 & 0.86\\
4350 & 3.27 $\pm$ 0.12 & 14.05-20.00 & 8 &-13.08 $\pm$ 0.03  & 14.05 $\pm$ 0.04 & 1.49&130.96  $\pm$  11.84 & -1.55 $\pm$ 0.08 & 0.56 $\pm$ 0.14 & 0.66 $\pm$ 0.17 & 1.10\\
5478 & 2.96  $\pm$ 0.12 & 2.08-12.00 & 11 & 0.52 $\pm$ 0.05 & 2.08 $\pm$ 0.04 &0.97&156.36   $\pm$ 4.63 &  -0.02 $\pm$ 0.09 & 0.70 $\pm$ 0.12 & 0.75 $\pm$ 0.08 & 1.58\\
5517 & 1.77  $\pm$ 0.11 & 0.83-6.00  & 9 & 0.18 $\pm$ 0.44  &0.83 $\pm$  0.07&0.97&157.15 $\pm$ 13.86 & -1.18 $\pm$ 0.13 & 0.73 $\pm$ 0.21 & 0.77 $\pm$ 0.21 & 0.62\\
5523 & 3.67  $\pm$ 0.14 & 1.05-5.00 & 5 & 0.33 $\pm$  0.18 &1.05 $\pm$ 0.03&1.16&171.85  $\pm$  11.44 & -1.01 $\pm$ 0.09 & 1.01 $\pm$ 0.22 & 1.10 $\pm$  0.25 & 0.67\\
5601 & 4.49  $\pm$ 0.14  & 7.67-15.00 & 25  & -5.04 $\pm$  0.18 &  7.67 $\pm$ 0.02&1.31& 205.55 $\pm$9.44  & -0.72 $\pm$ 0.06 & 1.06 $\pm$ 0.05 & 1.02 $\pm$ 0.11& 0.99\\
6159 & 1.94  $\pm$ 0.12 & 3.17-12.00 & 7 & 2.28 $\pm$ 8.25 & 3.17 $\pm$  0.11 & 1.01& 68.91  $\pm$ 5.12 & -0.09  $\pm$ 0.56 & 0.65 $\pm$ 0.17 & 0.70 $\pm$ 0.21& 0.89\\
6397 & 5.78  $\pm$ 0.15  & 3.42-25.00 &13 & 0.55 $\pm$ 0.05  &3.42 $\pm$  0.02& 1.40 &192.55 $\pm$ 5.44 & -0.65 $\pm$ 0.04& 0.55 $\pm$ 0.06 &0.64 $\pm$ 0.06 & 1.72\\
6504& 2.33  $\pm$ 0.12 & 3.09-17.00& 10 & 1.04 $\pm$ 0.51& 3.09 $\pm$ 0.07 & 0.98 &155.32  $\pm$ 4.29 & 0.43  $\pm$ 0.11 & 0.74 $\pm$ 0.08 & 0.77 $\pm$ 0.06 & 0.62 \\
6621 & 6.71 $\pm$ 0.16  & 32.53-37.00 & 10 & -29.56 $\pm$ 0.48 & 32.53 $\pm$ 0.02 &1.09 & 95.04  $\pm$ 4.87 &  -0.98 $\pm$   0.13 & 0.89 $\pm$ 0.23 & 0.81 $\pm$ 0.13 & 2.58\\
6625 & 1.81 $\pm$ 0.13  & 5.22-23.00  & 13 & 1.51 $\pm$ 0.84  & 5.22 $\pm$  0.11& 1.07 &82.74 $\pm$ 1.98 &  -0.89 $\pm$  0.11 & 0.79 $\pm$ 0.10 & 0.76 $\pm$ 0.11 & 0.86\\
6657 & 1.86 $\pm$ 0.21 & 4.26-47.00 & 5 &1.31 $\pm$  0.11 &4.26 $\pm$ 0.19 &1.06& 70.41  $\pm$  11.14 & 1.03 $\pm$ 1.38& 0.36 $\pm$ 0.12 & 0.41 $\pm$  0.09 & 0.17\\
6930 & 5.54 $\pm$ 0.18 & 31.83-35.00 & 9  & -29.94 $\pm$  0.15 &  31.83 $\pm$ 0.02&1.06 &95.18 $\pm$ 4.34  & -0.64 $\pm$ 0.13 & 1.12 $\pm$ 0.17 & 1.02 $\pm$ 0.16& 0.19\\
7293 & 2.95 $\pm$0.11  & 3.66-40.00 & 19 & 0.46 $\pm$ 0.05 & 3.66 $\pm$  0.06 & 1.27& 161.44 $\pm$ 3.26 & 0.32 $\pm$ 0.08 & 0.60 $\pm$ 0.02 & 0.50 $\pm$ 0.02& 0.99\\
7295 &  3.26  $\pm$ 0.17 & 2.23-9.00 &6 & 1.16 $\pm$ 0.64  &2.23 $\pm$  0.06& 1.02 &344.84 $\pm$ 14.25 & 0.49 $\pm$ 0.10 & 0.53 $\pm$ 0.12 &0.61 $\pm$ 0.13 & 0.72\\
7475 & 3.69  $\pm$ 0.14 & 9.15-30.00& 16 & -0.05 $\pm$ 0.11& 9.15 $\pm$ 0.05 & 1.84 &173.14 $\pm$ 12.34  & -1.29 $\pm$ 0.05 & 0.64 $\pm$ 0.04 & 0.69 $\pm$ 0.09 & 0.67 \\
7548 & 2.95 $\pm$ 0.12 & 3.77-6.00 & 5 & -1.64 $\pm$ 0.22&3.77 $\pm$ 0.03 &  0.98 & 217.14 $\pm$ 14.34 &  -0.91 $\pm$ 0.07 & 1.14 $\pm$ 0.71&1.05 $\pm$ 0.78 & 0.15\\
7588 & 2.08  $\pm$ 0.12 & 2.74-13.00 & 5 & 0.44  $\pm$  0.27 & 2.74  $\pm$  0.07 & 0.84 &  78.84 $\pm$ 2.35 & -0.68 $\pm$ 0.17 & 0.44 $\pm$ 0.08 & 0.54 $\pm$ 0.10 & 0.34\\
7638 & 1.75 $\pm$ 0.11 & 1.10-9.00 & 16& 0.92 $\pm$ 0.63 & 1.10 $\pm$ 0.05 & 1.47 & 61.34  $\pm$ 3.23 & -0.92 $\pm$ 0.51 & 0.32 $\pm$ 0.11 & 0.41 $\pm$ 0.11 & 0.34\\
7648 & 1.55 $\pm$ 0.10 & 4.43-31.00 & 10 & 0.89 $\pm$ 0.46 & 4.43 $\pm$ 0.14 & 1.02 & 252.93 $\pm$ 9.04 &  -0.44 $\pm$  0.06 & 0.58 $\pm$ 0.08 & 0.57 $\pm$ 0.08 & 0.40\\
7711 & 3.66 $\pm$ 0.13 & 1.98-8.00 & 10 & 2.01 $\pm$ 1.78 & 1.98 $\pm$ 0.04 & 1.28 & 211.03 $\pm$ 12.83 &  -1.24  $\pm$  0.05 & 1.06 $\pm$ 0.19 & 1.05 $\pm$ 0.18 & 0.98\\
8049$_{-}$1 & 1.59 $\pm$ 0.09 & 8.03-20.00 & 7 & 15.45 $\pm$ 9.02 & 8.03 $\pm$ 0.16 & 1.02 & 149.13  $\pm$ 5.45 & -0.32 $\pm$ 0.11 & 1.21 $\pm$ 0.55 & 1.02 $\pm$ 0.57 & 0.37 \\
8049$_{-}$2 & 1.59 $\pm$ 0.09 & 30.99-80.00 & 9 & 13.15 $\pm$ 10.17 & 30.99 $\pm$ 0.18 & 0.96 &  115.03 $\pm$ 4.38& -1.26 $\pm$ 0.08 & 1.17 $\pm$ 0.15 & 1.04  $\pm$ 0.17 & 1.68\\
\enddata
\tablecomments{$F_{p}$ denotes the peak flux on a 256 ms timescale;
time interval is the interval of start and end time of decay phase
of selected pulses.  $\chi_{\nu}^{2}1$ and $\chi_{\nu}^{2}2$ are the
fitting $\chi_{\nu}^{2}$ of pulse light curves and the power-law
decay of $E_{p}$, respectively.}
\end{deluxetable}

\begin{deluxetable}{ccccccccccccc}   %
\tabletypesize{\scriptsize}  \tablecaption{A list of burst sample
and various  parameters for sample 2.
\label{tbl-1}}\tablewidth{0pt}\rotate \tablehead{ \colhead{Trigger}
& \colhead{$F_{p}$} &\colhead{time interval} & \colhead{n}&
\colhead{$t_{0}$} & \colhead{$t_{m}$} & \colhead{$\chi_{\nu}^{2}1$}
&
\colhead{$E_{p}$}& \colhead{$\alpha$}& \colhead{$\delta$} & \colhead{S} & \colhead{$\chi_{\nu}^{2}2$}\\
&(phs cm$^{-2}$s$^{-1}$)& (s) & & (s) &(s)& & (keV)& & }\startdata
&  &  &&  & Band photon model &  & & & &  \\
\hline
\\
563 & 1.89 $\pm$ 0.14 & 1.84-32.00 & 12  &  0.66 $\pm$ 0.04  & 1.84 $\pm$ 0.05 & 1.25 & 169.53 $\pm$ 4.74 & -0.58 $\pm$ 0.07& 0.75 $\pm$ 0.04 & 0.78 $\pm$ 0.09 & 1.62 \\
907 &  3.57 $\pm$ 0.17 & 1.67-16.00 & 16 &  0.31 $\pm$ 0.04  & 1.67 $\pm$ 0.03 & 1.38 & 198.34 $\pm$ 3.30& 0.26 $\pm$ 0.06& 0.67 $\pm$ 0.02  &0.74 $\pm$ 0.01 & 1.61 \\
914 & 2.53  $\pm$ 0.16 &0.61-5.00 & 6 & 1.17 $\pm$ 0.45 & 0.61$\pm$ 0.02  & 0.92&116.46 $\pm$ 4.77&-1.29 $\pm$ 0.08 & 1.20 $\pm$ 0.21  & 1.21 $\pm$ 0.20 & 1.25 \\
973$_{-}$1 & 5.29 $\pm$ 0.20 & 2.79-22.00 & 21 & -0.41 $\pm$ 0.07  & 2.79 $\pm$ 0.03 & 1.34& 412.52 $\pm$ 14.11& -1.26 $\pm$ 0.02 & 0.95 $\pm$ 0.09 & 1.00 $\pm$ 0.06 & 1.41 \\
973$_{-}$2 & 5.29 $\pm$ 0.20 & 24.09-35.00 & 7 & -22.14 $\pm$ 0.01 & 24.09 $\pm$ 0.03 & 1.24& 296.21 $\pm$ 24.92 & -1.51  $\pm$ 0.04 & 0.82 $\pm$ 0.15 & 0.78 $\pm$ 0.14 & 1.25\\
999 & 11.55 $\pm$ 0.31 & 3.99-6.00 & 13 & -3.22 $\pm$ 0.14 & 3.99 $\pm$ 0.00 & 2.37& 600.93 $\pm$ 31.72& -1.12 $\pm$ 0.03 & 1.31 $\pm$ 0.31 & 1.34  $\pm$ 0.26 & 0.98 \\
1406 & 1.97 $\pm$ 0.13  & 3.35-37.00 & 14 & 0.95 $\pm$ 0.07 & 3.35 $\pm$ 0.05 & 1.07&153.71 $\pm$ 4.84 &  -1.26 $\pm$ 0.05 & 0.79 $\pm$ 0.09 & 0.81 $\pm$ 0.08 & 0.78\\
1467 & 2.26 $\pm$ 0.13 & 4.47-28.00 & 5 & 0.69 $\pm$ 0.52 & 4.47 $\pm$ 0.04 & 0.99 & 111.64 $\pm$ 3.73 & -1.01 $\pm$ 0.12 & 0.41 $\pm$ 0.04 & 0.44 $\pm$ 0.04 & 1.81 \\
1733 & 3.00 $\pm$  0.15 & 3.46-30.00 & 5 & 6.26 $\pm$ 3.25 & 3.46 $\pm$  0.04 & 1.41 & 649.54 $\pm$ 79.23 & -1.17 $\pm$ 0.04 & 1.56 $\pm$ 0.34 & 1.21 $\pm$ 0.17 & 1.66\\
1883 & 5.20 $\pm$ 0.18 &1.31-12.00 & 7 & 0.25 $\pm$ 0.07 & 1.31 $\pm$ 0.02 & 0.97& 248.32 $\pm$ 16.72 &-1.37 $\pm$ 0.04 & 0.68 $\pm$ 0.17 &0.74 $\pm$ 0.15 &  1.56 \\
1956 & 2.57 $\pm$ 0.13 & 2.91-15.00 &  8 &  0.07 $\pm$ 0.29& 2.91 $\pm$ 0.04& 1.21 &146.23 $\pm$ 6.24 & -1.33 $\pm$  0.06& 0.76 $\pm$ 0.14 & 0.69 $\pm$ 0.13 & 1.65\\
1989 & 2.73 $\pm$ 0.15 & 116.54-130.00 & 5 & -111.58 $\pm$ 0.20 & 116.54 $\pm$ 0.05 & 1.32&105.32 $\pm$ 3.29&  -1.33 $\pm$ 0.06 & 0.93 $\pm$ 0.05  & 1.00 $\pm$ 0.09 & 1.01\\
2083$_{-}$1 & 45.42  $\pm$ 0.46 & 1.05-6.00 & 17  & 0.36 $\pm$ 0.22& 1.04 $\pm$ 0.00& 5.98&404.25  $\pm$  6.28  &-1.06 $\pm$ 0.01 & 1.34 $\pm$ 0.09 & 1.41 $\pm$ 0.03 & 1.15\\
2083$_{-}$2 & 45.42  $\pm$ 0.46 & 8.68-20.00 & 42 & -4.02 $\pm$ 0.78 & 8.68 $\pm$ 0.01 & 2.26&64.31 $\pm$ 2.54&-1.64 $\pm$ 0.03 & 0.85 $\pm$ 0.34 & 0.78 $\pm$ 0.04 & 0.77 \\
2138 & 7.00  $\pm$ 0.20 & 1.29-15.00 & 17& 1.28 $\pm$ 0.21 & 1.29 $\pm$ 0.06 &1.17&205.92  $\pm$ 17.4&  -0.49  $\pm$   0.18 & 0.45 $\pm$ 0.02 & 0.58 $\pm$ 0.15 & 0.58\\
2193 & 1.55  $\pm$ 0.13 &10.32-42.00 & 15 &-0.04 $\pm$ 0.5 & 10.32 $\pm$ 0.24 & 0.98 &295.31 $\pm$ 15.18 & 0.23  $\pm$ 0.07& 0.75 $\pm$ 0.06 & 0.96 $\pm$ 0.04 & 1.25\\
2387 & 3.86  $\pm$ 0.16 & 6.49-40.00 & 15 & -0.53 $\pm$ 0.27 &6.49 $\pm$ 0.07 &1.15&167.22 $\pm$ 3.13 &-0.75 $\pm$ 0.04 & 0.75 $\pm$ 0.03 & 0.77 $\pm$ 0.03 & 2.48\\
2484 & 1.55  $\pm$ 0.13 & 2.01-30.00 & 8 & 2.66 $\pm$ 0.85 & 2.01 $\pm$ 0.07 &  1.09&106.55 $\pm$   1.65 & -0.72 $\pm$ 0.03& 1.05 $\pm$ 0.13 & 0.97 $\pm$ 0.11 & 0.99\\
2519 & 1.53  $\pm$ 0.14 & 0.59-15.00 & 7& 0.45 $\pm$ 0.02 &0.59 $\pm$ 0.06 & 1.02 &192.76 $\pm$ 14.73 &  -0.97 $\pm$  0.12& 0.75 $\pm$ 0.07 & 0.75 $\pm$ 0.07 & 1.00\\
2662 & 1.52  $\pm$ 0.14 & 1.31-12.00 & 6 & 0.30 $\pm$ 0.05 &1.31 $\pm$ 0.07 &1.11&153.14  $\pm$  7.99 & -0.93 $\pm$ 0.12 &0.67 $\pm$ 0.07 & 0.69 $\pm$ 0.07 & 1.45\\
2665 & 1.99  $\pm$ 0.15 & 1.42-8.00  & 8 & 1.22 $\pm$ 0.32  &1.42 $\pm$ 0.05 & 0.89&106.03 $\pm$  2.24 &  -0.19 $\pm$ 0.11 & 0.75 $\pm$ 0.12 & 0.79 $\pm$ 0.25 & 1.21\\
2700 & 4.06  $\pm$ 0.18 & 53.7-61.00 & 19 & -52.84 $\pm$ 0.01 & 53.71 $\pm$ 0.03 &1.37&255.36  $\pm$  11.13 &  -1.35 $\pm$  0.03 & 0.45 $\pm$ 0.21 & 0.50 $\pm$ 0.14 & 0.42\\
2880 & 2.90  $\pm$ 0.14 &0.47-3.00  & 6 & 0.34 $\pm$ 0.04 &0.47 $\pm$ 0.01&1.43&125.93 $\pm$ 5.46 & -0.77 $\pm$  0.14 &0.76 $\pm$ 0.12 & 0.70 $\pm$ 0.14 & 1.10\\
2919 & 5.77  $\pm$ 0.19 &0.34-7.00 & 17 & 1.57  $\pm$ 0.12 &0.34 $\pm$ 0.02 & 1.39&477.43 $\pm$ 26.64 & -1.34 $\pm$  0.04 & 0.70 $\pm$ 0.11& 0.69 $\pm$ 0.11 & 0.62\\
3003 & 2.83 $\pm$ 0.16  & 9.68-24.00 & 17 & 8.94 $\pm$ 7.15 &9.68  $\pm$ 0.08 &1.02&473.83  $\pm$ 39.82 & -1.22  $\pm$ 0.04 & 0.85 $\pm$ 0.06 & 1.03 $\pm$ 0.43 & 1.20\\
3143 & 2.59  $\pm$ 0.14 & 0.69-4.00 & 6 & 0.64 $\pm$ 0.25 & 0.69 $\pm$ 0.02 & 0.96&216.09  $\pm$  30.27 & -1.53 $\pm$  0.09 & 1.34 $\pm$ 0.44 & 1.30 $\pm$ 0.43 & 1.03\\
3256 & 1.76  $\pm$ 0.11 & 1.38-9.00 & 27 & 0.55 $\pm$ 0.08 &1.38 $\pm$ 0.07& 1.06 &151.93   $\pm$  5.39 & -0.50 $\pm$ 0.15 & 0.65 $\pm$ 0.05 & 0.68 $\pm$ 0.06 & 0.91\\
3257 & 3.06  $\pm$ 0.13 & 3.51-35.00 & 29 & 0.60 $\pm$ 0.06 & 3.51 $\pm$ 0.06 & 1.10&203.33 $\pm$ 5.26 &  -0.31 $\pm$ 0.05& 0.45 $\pm$ 0.02 & 0.45 $\pm$ 0.03 &2.05\\
3290  & 10.71  $\pm$ 0.18 & 2.98-4.00 & 5& -2.73 $\pm$ 0.00 &2.98 $\pm$ 0.01& 2.52& 38.45 $\pm$ 13.53 & -1.86 $\pm$ 0.87 & 0.72 $\pm$ 0.30 & 0.69 $\pm$ 0.26 & 1.19\\
3415 & 9.16  $\pm$ 0.19 & 11.56-15.00  & 6 & -10.84 $\pm$ 0.02 & 11.56 $\pm$ 0.01&1.44&426.74 $\pm$ 30.54 & -1.31  $\pm$ 0.04 & 0.75 $\pm$ 0.23 & 0.72 $\pm$ 0.19 & 1.78\\
3648$_{-}$1 & 5.70  $\pm$ 0.15 & 2.72-14.00 & 8 & 2.05 $\pm$ 0.00 & 2.72 $\pm$ 0.15& 0.81&100.22  $\pm$  1.68  & 0.32 $\pm$ 0.15 & 0.49 $\pm$ 0.04 & 0.50 $\pm$ 0.03 & 0.74\\
3648$_{-}$2 & 5.70  $\pm$ 0.15 & 23.85-36.00 & 6 & -19.47 $\pm$ 0.19 & 23.85 $\pm$ 0.18& 0.97&94.09  $\pm$  2.28  &-0.51 $\pm$ 0.15 & 0.48 $\pm$ 0.06 & 0.46 $\pm$ 0.05 & 1.16\\
3648$_{-}$3 & 5.70  $\pm$ 0.15  & 41.04-48.00 & 12 & -32.20 $\pm$ 1.44 & 41.04 $\pm$ 0.02 & 1.36&197.94 $\pm$ 4.30&-0.95 $\pm$ 0.05& 0.69 $\pm$ 0.02 & 0.75 $\pm$ 0.09 & 1.85\\
3765 & 25.29  $\pm$ 0.27 &66.15-73.00 & 12 &-63.70 $\pm$ 0.24 & 66.15 $\pm$ 0.01 & 1.45&335.73 $\pm$ 2.94 &  -0.93 $\pm$  0.15 & 1.05 $\pm$ 0.19 &1.04 $\pm$ 0.05 & 0.77\\
3870 & 13.90  $\pm$ 0.23 & 0.50-6.00 & 21  & 0.46 $\pm$ 0.38  & 0.50 $\pm$ 0.01& 1.32&313.03  $\pm$ 39.34 & -1.82 $\pm$ 0.02& 1.08 $\pm$ 0.11  & 1.00 $\pm$ 0.10 & 1.16\\
3954 &  8.19 $\pm$ 0.19  & 0.78-9.00 & 8 & 2.26 $\pm$ 1.47 & 0.78 $\pm$ 0.01 & 1.05&520.15  $\pm$ 35.34 & -1.3 $\pm$ 0.03& 0.77 $\pm$  0.13 & 0.78 $\pm$ 0.14 &1.41\\
4350 & 3.27  $\pm$ 0.12 & 14.05-20.00 & 10 &-13.08 $\pm$ 0.03  & 14.05 $\pm$ 0.04 & 1.49&79.96  $\pm$  11.84 & -1.55 $\pm$   0.12 & 0.69 $\pm$ 0.14 & 0.78 $\pm$ 0.17 & 0.45\\
5478 & 2.96  $\pm$ 0.12 & 2.08-12.00 & 20 & 0.52 $\pm$ 0.05 & 2.08 $\pm$ 0.04 &0.97&162.36 $\pm$ 3.63 &  -0.11 $\pm$ 0.08 & 0.63 $\pm$ 0.12 & 0.67 $\pm$ 0.08 & 1.48\\
5517 & 1.77  $\pm$ 0.11 & 0.83-6.00  & 12 & 0.18 $\pm$ 0.44  &0.83 $\pm$  0.07&0.97&157.15 $\pm$ 12.86 & -1.18 $\pm$ 0.13 & 0.97 $\pm$ 0.21 & 0.98 $\pm$ 0.21 & 0.85\\
5523 & 3.67  $\pm$ 0.14 & 1.05-5.00 & 5 & 0.33 $\pm$  0.18 &1.05 $\pm$ 0.03&1.16&188.85  $\pm$  12.44 & -1.11 $\pm$ 0.06 & 0.73 $\pm$ 0.12 & 0.78 $\pm$  0.25 & 0.35\\
5601 & 4.49  $\pm$ 0.14  & 7.67-15.00 & 14  & -5.04 $\pm$  0.18 &  7.67 $\pm$ 0.02&1.31&259.55  $\pm$ 7.44  & -0.94 $\pm$ 0.03 & 0.95 $\pm$ 0.05 & 1.04 $\pm$ 0.11& 1.26\\
6159 & 1.94  $\pm$ 0.12 & 3.17-10.00 & 5 & 2.28 $\pm$ 8.25 & 3.17$\pm$  0.11 & 1.01& 95.91  $\pm$ 3.12 & -1.61  $\pm$ 0.06 & 0.94 $\pm$ 0.31 & 1.01 $\pm$ 0.31& 0.89\\
6397 & 5.78  $\pm$ 0.15  & 3.42-40.00 &18 & 0.55 $\pm$ 0.05  &3.42 $\pm$  0.02& 1.40 &211.55 $\pm$ 4.44 & -0.75 $\pm$ 0.03& 0.65 $\pm$ 0.06 &0.70 $\pm$ 0.06 & 1.12\\
6504& 2.33  $\pm$ 0.12 & 3.09-50.00& 23 & 1.04 $\pm$ 0.51& 3.09 $\pm$ 0.07 & 0.98 &169.82  $\pm$ 4.11 & -0.74  $\pm$ 0.03 & 0.77 $\pm$ 0.08 & 0.73 $\pm$ 0.06 & 2.12 \\
6621 & 6.71 $\pm$ 0.16  & 32.53-37.00 & 11 & -29.56 $\pm$ 0.48 & 32.53 $\pm$ 0.02 &1.09 & 118.04  $\pm$ 3.87 &  -1.39 $\pm$   0.03 & 0.98 $\pm$ 0.23 & 0.96 $\pm$ 0.13 & 1.41\\
6625 & 1.81 $\pm$ 0.13  & 5.22-25.00  & 21 & 1.51 $\pm$ 0.84  & 5.22 $\pm$  0.11& 1.07 &87.74 $\pm$ 1.98 &  -1.39$\pm$  0.04 & 0.65 $\pm$ 0.10 & 0.63 $\pm$ 0.11 & 1.08\\
6657 & 1.86 $\pm$ 0.21 & 4.26-57.00 & 13 &1.31 $\pm$  0.11 &4.26 $\pm$ 0.19 &1.06& 122.41  $\pm$  9.14 &   -1.06 $\pm$ 0.17& 0.25 $\pm$ 0.12 & 0.43 $\pm$  0.09 & 0.97\\
6930 & 5.54 $\pm$ 0.18 & 31.83-35.00 & 9  & -29.94 $\pm$  0.15 &  31.83$\pm$ 0.02&1.06 &127.18 $\pm$ 3.34  & -1.17 $\pm$ 0.04 & 1.07 $\pm$ 0.17 & 1.05 $\pm$ 0.16& 1.01\\
7293 & 2.95 $\pm$ 0.11  & 3.66-40.00 & 19 & 0.46 $\pm$ 0.05 & 3.66 $\pm$  0.06 & 1.27& 184.44 $\pm$ 2.26   & 0.29   $\pm$ 0.04 & 0.54 $\pm$ 0.02 & 0.50 $\pm$ 0.02& 0.94\\
7295 & 3.26 $\pm$ 0.17 & 2.23-9.00 &7 & 1.16 $\pm$ 0.64  &2.23 $\pm$  0.06& 1.02 &368.84 $\pm$ 10.25 & 0.23 $\pm$ 0.06 & 0.52 $\pm$ 0.12 &0.53 $\pm$ 0.13 & 2.52\\
7475 & 3.69  $\pm$ 0.14 & 9.15-30.00& 17 & -0.05$\pm$ 0.11& 9.15 $\pm$ 0.05 & 1.84 &279.14 $\pm$ 10.34  & -1.49 $\pm$ 0.04 & 0.64 $\pm$ 0.04 & 0.68 $\pm$ 0.09 & 1.56 \\
7548 & 2.95 $\pm$ 0.12 & 3.77-20.00 & 7 & -1.64 $\pm$ 0.22&3.77 $\pm$ 0.03 &  0.98 & 197.84 $\pm$ 14.34 &  -0.87 $\pm$ 0.06 & 0.66 $\pm$ 0.15&0.76 $\pm$ 0.08 & 2.15\\
7588 & 2.08  $\pm$ 0.12 & 2.74-13.00 & 5 & 0.44  $\pm$  0.27 & 2.74  $\pm$  0.07 & 0.84 &  84.87 $\pm$ 2.15 & -1.08 $\pm$ 0.07 & 0.54 $\pm$ 0.08 & 0.55 $\pm$ 0.18 & 0.39\\
7638 & 1.75 $\pm$0.11 & 1.10-9.00 & 16& 0.92 $\pm$ 0.63 & 1.10 $\pm$ 0.05 & 1.47 & 58.44  $\pm$ 3.23 & -1.86 $\pm$ 0.09 & 0.59 $\pm$ 0.12 & 0.61 $\pm$ 0.11 & 0.34\\
7648 & 1.55 $\pm$ 0.10 & 4.43-31.00 & 14 & 0.89 $\pm$ 0.46 & 4.43 $\pm$ 0.14 & 1.02 & 252.93 $\pm$ 8.42 &  -0.44 $\pm$  0.06 & 0.55 $\pm$ 0.05 & 0.66 $\pm$ 0.06 & 0.58\\
7711 & 3.66 $\pm$ 0.13 & 1.98-8.00 & 23 & 2.01 $\pm$ 1.78 & 1.98 $\pm$ 0.04 & 1.28 & 216.03 $\pm$ 9.99 &  -1.26 $\pm$ 0.04 & 1.29 $\pm$ 0.09 & 1.28 $\pm$ 0.08 & 1.14\\
8049$_{-}$1 & 1.59 $\pm$ 0.09 & 8.03-20.00 & 17 & 15.45 $\pm$ 9.02 & 8.03 $\pm$ 0.16 & 1.02 & 150.53  $\pm$ 5.45 & -0.34 $\pm$ 0.11 & 1.18 $\pm$ 0.15 & 1.38 $\pm$ 0.16 & 0.92 \\
8049$_{-}$2 & 1.59 $\pm$ 0.09 & 30.99-80.00 & 26 & 13.15 $\pm$ 10.17 & 30.99 $\pm$ 0.18 & 0.96 &  115.03 $\pm$ 4.38 & -1.26 $\pm$ 0.01& 1.26 $\pm$ 0.12 & 1.16  $\pm$ 0.11 & 0.69\\
8111 & 4.21 $\pm$ 0.14 & 4.98-11.00& 10 & -4.22 $\pm$ 0.03 & 4.98 $\pm$ 0.02 & 1.17 & 253.93 $\pm$ 22.51 & -1.54 $\pm$ 0.04 & 0.71 $\pm$ 0.16& 0.76 $\pm$ 0.11 & 1.06\\
\enddata
\tablecomments{the meanings of the parameters are the same as table
1.}
\end{deluxetable}

\section{conclusions and discussions}
Pulses are the basic, central building blocks of GRB prompt
emission, and it is essential to our understanding of GRB physics
(Hakkila et al. 2008). In this paper we select a well-separated
single pulses presented by Kocevski et al. (2003) and Norris et al.
(1999) including bright and weak bursts to investigate the
evolutionary slope of $E_{p}$ during the pulse decay phase. We first
fit the $E_{p}$ of time-resolved spectra for our selected pulses
with a single power-law function form and obtain the evolutionary
slope to compare it with the theoretical predication of Paper I. Our
results show that these observed evolutionary slopes are in good
agreement with the predictions of Paper I. Contrary to Paper I,
however, we find that within the limits of uncertainty the
corresponding intrinsic spectra of most of bursts (about 25 for
sample 1 and 29 for sample 2) may bear the intrinsic Comptonized or
thermal synchrotron spectrum, whereas only a small number of bursts
(about 16 for sample 1 and 12 for sample 2) may be associated with
the intrinsic Band function spectra. Paper I investigated 12 pulses
and found that there are eight consisting with intrinsic Band
spectrum and only two corresponding to the intrinsic Comptonized or
thermal synchrotron spectrum. We argue our analysis results are more
reliable for the following reasons. Firstly, the quantity of our
sample (56 for Band model and 59 for Compton model) much larger than
that of Paper I. Secondly, we remove all possible
improperly-measured $E_{p}$, i.e. the $E_{p}$ with $\alpha < -2$ and
$\beta > -2$ for the BAND model, $\alpha < -2$ for the COMP model
are discarded, but Paper I does not seem to consider it. Lastly, the
$E_{p}$ with its error larger than 50 percent of itself are also
excluded but Paper I includes the $E_{p}$ with much larger error.
Since the Paper I pointed out that the evolutionary curve of the
decay portion of the pulse is dominated by the curvature effect our
results also indicate that most of the FRED pulses we selected
indeed result from this effect.

There are also some outliers (about 15 for sample 1 and 18 for
sample 2) included in our sample. Besides the curvature effect,
there might be some other factors that can affect the value of S.
One factor would be the variation of the rest-frame emission
mechanism, which was revealed in Qin et al. (2005). For example,
different rest-frame spectra or different speeds of the rest-frame
spectral softening could lead to different values of the power-law
index. Other factors pointed out by Paper I, such as the real
intrinsic radiation mechanism may be more complicated as well as
GRBs might be associated with more complicated situations rather
than a simple fireball expanding isotropically with a constant
Lorentz factor. There are some disparate $E_p$ values that could be
the result of source confusion (pulse overlapping) in the pulse
decay phase (e.g. we can see from Figures 1 and 2). Maybe this is
also one of the reasons giving arise to the outliers.

We also study the possible correlations between the S and other
parameters. A strong correlation between S and the $E_{p,max}$
during the phase of pulses decay is identified. There might be some
important link between S and the $E_{p,max}$. Also an important
correlation between the observed $E_{p}$ of pulse time-integrated
spectra and S exists. In order to reveal if this correlation is
intrinsic we attempt to use the pseudo-redshift calculated by
Yonetoku et al. (2004) to check it. We also find there is
correlation between the S and intrinsic peak energy $E_{p,i}$ but
for sample 2 the intrinsic correlation is looser than the observed
one. One of reasons we consider is that the pseudo-redshifts may not
well represent the real redshifts of our selected bursts. Another
reason we suspect that the Band model may be more suitable for
describing the GRB time-integrated spectra than Compton model.

$E_{p}$ is one of the fundamental characteristics of the prompt
emission of GRB. Based on it several important empirical relations
have been proposed. One of important relations is the so-called
Amati relation (Amati et al. 2002; Amati 2003), i.e. the correlation
between $E_{p}$ in the GRB rest frame ($E_{p}^{src}$) and the
isotropic radiated energy ($E^{iso}$). The second correlation with
much tighter correlation than Amati relation is between the
$E_{p}^{src}$ energy and the collimation-corrected energy
($E_{\gamma}$), the so-called Ghirlanda relation (Ghirlanda et al.
2004). Similarly Liang \& Zhang (2005) found a good correlation
between $E_{p}^{src}$, $E^{iso}$, and the achromatic break time in
the afterglow light curve ($t_{jet}$). The third relationship is
between $E_{p}^{src}$ and the isotropic peak luminosity ($L^{iso}$),
the so-called the $E_{p}^{src}$, $L^{iso}$ (Yonetoku) relation
(Yonetoku et al. 2004). If these relations are valid, they must be
related to the fundamental physics of GRBs. Thus, $E_{p}^{src}$
energy provides us important knowledge about the characteristics of
the prompt emission of GRBs. In order to calculate the bolometric
fluence which reflects the total radiated energy in the prompt
emission the observed $E_{p}$ is a crucial quantity. Therefore, the
evolutionary slope might be a useful quantity to help us understand
the physic of prompt emission since the evolutionary slope is
related to the observed $E_{p}$ as well as the intrinsic $E_{p}$.

The observed $E_p$ decay is clearly related to the pulse peak lag
described by Hakkila et al. (2008) and Hakkila \& Cumbee (2009).
Both the $E_p$ decay and the pulse peak lag  describe pulse spectral
softening from different viewpoints (the $E_p$ decay reveals
spectral changes with time, while pulse peak lag examines a
self-similar temporal structure at different energies). There are
several interesting and pertinent repercussions of this.

Firstly, pulse evolution is not limited to the ``pulse decay''
phase. Although the pulse peak intensity is defined to be the time
at which the decay begins, this is not the case: pulse peak lags
indicate that the low-energy count rate is still increasing at the
time of pulse peak intensity whereas the high-energy count rate has
already started to decay. So, when does the pulse $E_p$ decay begin
if not at the time of pulse peak intensity? The evidence presented
here from the plots in Figures 1 and 2 suggests that for most of
pulses it probably starts dropping at the pulse onset: the entire
pulse likely represents a decay of $E_p$ (this result has been
mentioned before by various other authors, e.g. Norris et al. 1986).
However, this observational result contradicts the theoretical
conclusion of Lu et al. (2006), which states, ``Now it is clear
that, in the mechanism of the curvature effect, it is the decaying
phase of a local pulse that contributes to the observed lags.''
Since Kocevski \& Liang (2003) argued that the observed lag is the
direct result of spectral evolution (the $E_p$ decay through the
four BATSE channels), something appears to be missing from the Lu et
al. (2006) theoretical models, based on observations? We shall
analyze it in detail in the future work. However, there are also
some pulses whose spectra exhibit ``tracking'' (the observed
spectral parameters follow the same pattern as the flux or count
rate time profile) behaviors (e.g. Crider et al. 1997; Peng et al.
2009). The trigger 5601 demonstrated in Figure 1 is such a
``tracking'' pulse. In this case, the $E_p$ starts dropping at
around the pulse peak intensity, while the $E_p$ rises during the
pulse rise phase. In fact the Hakkila et al. (2008) and Hakkila \&
Cumbee (2009) results have showed that spectral and intensity
evolution simply scale with the pulse duration. Short, ``tracking''
pulses should thus exhibit the same type of $E_p$ decay as the
longer pulses, but vary on too short a timescale for this variation
to have been recognized in many studies.

Secondly, pulse evolution appears to be a defining characteristic of
GRB pulses. The $E_p$ decline correlates with a wide range of other
pulse properties. Hakkila et al. (2008) showed that pulse duration
is tightly correlated with pulse lag and anti-correlated with pulse
peak luminosity, and it certainly appears that the same is true here
(for example, the $E_p$ decay correlates with the pulse decay time
interval). Similar to what is found in this work Hakkila \& Cumbee
(2009) also found that pulse peak lag anti-correlates with pulse
peak flux and correlates with pulse asymmetry, which reveals pulse
shape seems related to the pulse decay rate. It seems to mean that
pulse evolution is more than a characteristic of the pulse; it
appears to DEFINE the pulse.

Lastly, pulse evolution appears to drive the evolution observed in
GRB bulk emission. Most of GRBs exhibit hard-to-soft evolution, this
is due to the smeared measurement of overlapping pulses, and to the
fact that hard, short, luminous pulses typically occur near the
beginning of a burst. Our paper demonstrates that spectral evolution
exists at the even finer level of pulse evolution. The fact that
pulse properties are correlated (pulse peak lag dictates pulse
luminosity, duration, hardness, and asymmetry) suggests that
internal pulse evolution is responsible for pulse-to-pulse
variations and ultimately for the burst evolution.

In this study the pulses are limited to FRED pulses, which is based
on the fact that the single pulses often been described by a fast
rise and exponential decay as pointed out previously. Recently,
Hakkila et al. (2008) and Hakkila \& Cumbee (2009) suggested that
most pulses in long GRBs are really just variations of the same
phenomenon. Therefore, our analysis results about evolutionary slope
should apply to those non-FRED pulses.



\section{Acknowledgments}
We thank the anonymous referee for constructive suggestions. This
work was supported by the Natural Science Fund for Young Scholars of
Yunnan Normal University (2008Z016), National Natural Science
Foundation of China (No. 10778726), the Natural Science Fund of
Yunnan Province (2006A0027M).


\clearpage
\end{document}